\documentclass{article}
\usepackage{ijcai23}

\usepackage{times}
\usepackage{soul}
\usepackage{url}
\usepackage[utf8]{inputenc}
\usepackage[small]{caption}
\usepackage{graphicx}
\usepackage{amsmath}
\usepackage{amsthm}
\usepackage{booktabs}
\usepackage{algorithm}
\usepackage{algorithmic}
\usepackage{sectsty}
\usepackage[switch,mathlines]{lineno}

\nolinenumbers
\usepackage{microtype}
\usepackage{amssymb}
\usepackage{tikz}
\usetikzlibrary{patterns,decorations.pathreplacing, arrows.meta}
\usepackage{hyperref}
\hypersetup{
pdfencoding=auto, psdextra,
colorlinks=true,citecolor=green!30!black,linkcolor=red!40!black,urlcolor=blue!60!black
}
\usepackage[sort&compress,nameinlink]{cleveref}
\usepackage{nicefrac}
\usepackage{natbib}
\usepackage{comment}
\usepackage{paralist}
\usepackage{enumitem}
\usepackage{xcolor}
\usepackage{colortbl}
\usepackage{subcaption}
\usepackage{booktabs}
\usepackage{listings}
\usepackage[linecolor=green!70!black, backgroundcolor=green!10, bordercolor=black, textsize=small]{todonotes}
\definecolor{winered}{rgb}{0.5,0.1,0.1}

\renewcommand*{\le}{\leqslant}
\renewcommand*{\leq}{\leqslant}

\renewcommand*{\geq}{\geqslant}
\renewcommand{\epsilon}{\varepsilon}

\crefname{table}{Table}{Tables}
\Crefname{table}{Table}{Tables}
\crefname{figure}{Figure}{Figures}
\crefname{theorem}{Theorem}{Theorems}
\crefname{definition}{Definition}{Definitions}
\crefname{corollary}{Corollary}{Corollaries}
\crefname{observation}{Observation}{Observations}
\crefname{question}{Question}{Question}
\crefname{lemma}{Lemma}{Lemmas}
\crefname{example}{Example}{Examples}
\crefname{reduction}{Reduction}{Reductions}
\crefname{construction}{Construction}{Constructions}
\crefname{subsection}{Section}{Sections}
\crefname{section}{Section}{Sections}
\crefname{proposition}{Proposition}{Propositions}
\crefname{algorithm}{Algorithm}{Algorithms}
\crefname{algocf}{Algorithm}{Algorithms}
\Crefname{equation}{Inequality}{Inequalities}
\crefname{lstlisting}{listing}{listings}

\newcommand{\myemph}[1]{{\color{winered}\emph{#1}}}
\newcommand{\naturals}{{{\mathbb{N}}}}

\newcommand{\reals}{{{\mathbb{R}}}}
\newcommand{\pabulib}{{{\textsc{Pabulib}}}}
\newcommand{\pabutools}{{{\textsc{Pabutools}}}}
\newcommand{\pabustats}{{{\textsc{Pabustats}}}}
\newcommand{\cost}{{{\mathrm{cost}}}}

\renewcommand{\sc}{{{\mathrm{sc}}}}
\newcommand{\es}{{{\mathrm{ES}}}}

\renewcommand{\part}{{{\mathrm{part}}}}

\theoremstyle{definition}

\theoremstyle{plain}

\makeatletter
\newtheorem*{rep@theorem}{\rep@title}
\newcommand{\newreptheorem}[2]{%
\newenvironment{rep#1}[1]{%
 \def\rep@title{#2 \ref{##1}}%
 \begin{rep@theorem}}%
 {\end{rep@theorem}}}
\makeatother
\newreptheorem{theorem}{Theorem}
\newreptheorem{proposition}{Proposition}
\newreptheorem{lemma}{Lemma}
\newreptheorem{corollary}{Corollary}

\usepackage{soul}

\usepackage{verbatimbox}
\usepackage{verbatim}
\usepackage{verbments}

\newcommand{\bftt}[1]{\textbf{\texttt{#1}}}

\title{Participatory Budgeting: Data, Tools, and Analysis}

\author{
Piotr Faliszewski$^1$
\and
Jarosław Flis$^2$\and
Dominik Peters$^3$ \and
Grzegorz Pierczyński$^4$\and 
\\
Piotr Skowron$^4$\and
Dariusz Stolicki$^2$\and
Stanisław Szufa$^{1,2}$\And
Nimrod Talmon$^5$
\affiliations
$^1$AGH University,
$^2$Jagiellonian University in Krakow \\
$^3$CNRS, LAMSADE, Universit\'e Paris Dauphine - PSL,
$^4$University of Warsaw,
$^5$Ben-Gurion University
\emails
faliszew@agh.edu.pl,
jaroslaw.flis@uj.edu.pl,
mail@dominik-peters.de,
g.pierczynski@mimuw.edu.pl,
p.skowron@mimuw.edu.pl,
dariusz.stolicki@uj.edu.pl,
stanislaw.szufa@uj.edu.pl,
talmonn@bgu.ac.il
}

\begin{document}

\maketitle

\begin{abstract}
    We provide a library of participatory budgeting data (Pabulib) and open source tools (Pabutools and Pabustats) for analysing this data.
    We analyse how the results of participatory budgeting elections would change if a different selection rule was applied. 
    We provide evidence that the outcomes of the Method of Equal Shares would be considerably fairer than those of the Utilitarian Greedy rule that is currently in use. We also show that the division of the projects into districts and/or categories can in many cases be avoided when using proportional rules. We find that this would increase the overall utility of the voters. 
\end{abstract}

\section{Introduction}

Participatory budgeting (PB)~\citep{participatoryBudgeting} is a form of public consultation in which residents decide how to spend a part of the municipal budget. First, a number of projects are submitted, and after the initial formal evaluation, some of these projects are admitted to voting. Next, each citizen can participate in an election, and express their preferences over the projects. Finally, given the voters' ballots, a decision is made as to which of the projects to fund.

Each step in this process must be carefully designed in order to ensure that the selected projects match the voters' preferences to the highest possible degree. For example, some cities regulate the submission process by putting upper bounds on the costs of the projects (e.g., to avoid situations where a single expensive project consumes the whole budget, leaving a large fraction of the voters unsatisfied) or by assigning each project to one of a few predefined categories (e.g., to ensure that projects from the less popular categories also receive some funding). 
Choosing the right format of the ballots for the election also is a significant and challenging issue~\citep{benade2017preference}. For example, some cities use approval ballots, where the voters simply indicate which projects they support (sometimes with additional constraints, such as regarding the number of projects a voter can approve), while others turn to (forms of) score ballots, which allow the voters to indicate the degree of support for the respective projects. Further, some cities allow the voters to vote only on the projects from the district where they live, while others do not put such restrictions~\citep{her-kah-pet-pro:district-fair-pb}.

Finally, the voting rule used for aggregating the ballots and selecting the projects
is of utmost importance for the whole process.
For example, if the rule is capable of representing the voters proportionally, then it
may not be necessary to partition the projects into categories or to put constraints 
on their costs, whereas a majority-driven rule may indeed require such interventions.
Consequently, there is a growing interest in the design and analysis of voting rules for participatory budgeting~\citep{peters2021proportional, knapsackVoting, aziz2020participatory, talmon2019framework, aziz2018proportionally, fain2018fair, jiang2019approx, mun-she-wan-wan:approximate-core, sko-sli-szu-tal:pb-cumulative}. However, 
with a notable exception of \citet{ben-fai-gal:pb-real-experiments} (who performed lab experiments assessing how humans perceive different ballot formats),
all this research is focused on theoretical analysis. 

In this paper we take a step towards understanding how various selection rules for participatory budgeting operate in practice. We do this by releasing and analysing data from over 650 PB elections, mainly conducted in Poland.\footnote{Poland is a good source of PB instances because the law requires every major city to spend at least 0.5\% of its annual budget through PB. In 2021, over 42\% of Polish cities with populations above 5 000 organised PB elections, spending 627.5 million PLN in total.} Our contribution is the following: 


\begin{description}
\item[\pabulib.] This is a library of participatory budgeting data (PArticipatory BUdgeting LIBrary), and can be accessed via the following URL: \url{http://pabulib.org}.%
\item[\pabutools.] This is a Python library providing a parser of Pabulib files and implementations of selected rules for participatory budgeting. The  library is accessible via PyPI (\url{https://pypi.org/project/pabutools/}).
\item[\pabustats.] This is a web application for comparing various PB rules based on the data from Pabulib. It also offers the possibility to run simulations on files uploaded by the users. The application is accessible via the following URL: \url{http://pabulib.org/pabustats}.
\end{description}
We apply our tools on the collected data and perform an extensive analysis that compares different voting rules for PB.

\section{Preliminaries}

An \myemph{election} is a tuple $E = (P, N, b, \cost)$, where $P = \{p_1, \ldots, p_m\}$ is a set of \myemph{projects}, $N=\{1, 2, \ldots n\}$ is a set of \myemph{voters}, $b\in \naturals$ is the \myemph{budget}, and $\cost\colon P \to \naturals$ is a function that associates each project with its \myemph{cost}. The cost function naturally extends to sets: for each $W \subseteq P$, we let $\cost(W) = \sum_{p \in W} \cost(p)$. The voters express their preferences by casting \myemph{ballots}: each voter $i \in N$ assigns to each project $p \in P$ a score $s_i(p) \in \naturals$ that reflects her level of support for $p$. If $s_i(p) \in \{0, 1\}$ for all $i \in N$ and $p \in P$, then we have an \myemph{approval election}. Otherwise, we say it is a \myemph{cardinal election}. Intuitively, in an approval election each voter simply indicates the projects that she supports; in cardinal elections the voters provide more fine-grained information on how much they supports particular projects.


\subsection{Utility Models}

Given the ballot of voter $i \in N$, there are two natural ways to define $i$'s utility from a given set of projects $W \subseteq P$. The first approach is to assume that the utility does not depend on the costs of the projects; in this case we speak of \myemph{score utilities}, defined as $u^{\sc}_{i}(W) = \sum_{p \in W} s_i(p)$. For approval elections, this is the 
number of projects in $W$ that the voter supports.

An alternative approach is to assume that expensive projects carry more value to the voters; then, we speak of \myemph{cost utilities}, $u^{\cost}_{i}(W) = \sum_{p \in W} s_i(p)\cost(p)$. For approval elections, cost utilities can be interpreted as the amount of funds spent on the projects supported by a given voter.

Sometimes it will be clear from the context whether we refer to score or to cost utilities. In such cases, we will often omit the superscripts and write $u_i(W)$.

\subsection{Voting Rules}

A \myemph{voting rule} is a function that takes an election as input and returns a subset of projects $W$, called an \myemph{outcome}, such that $\cost(W) \leq b$. We say that outcome $W$ is \myemph{exhaustive} if for each project $c\in C\setminus W$ we have that $\cost(W) + \cost(c) > b$ (i.e., no additional project can be funded without violating the budget constraint). A voting rule $f$ is exhaustive if it always returns an exhaustive outcome.

Below we discuss two voting rules, Utilitarian Greedy and Method of Equal Shares, that we focus on in this paper. Each of them has two variants, depending on the type of utilities.

\paragraph{Utilitarian Greedy (UG).} We start with an empty outcome $W=\emptyset$, and 
repeatedly select a project $p$ maximising the ratio $\sum_{i \in N}u_i(p) / \cost(p)$. 
If $\cost(W) + \cost(p) \leq b$ then we add project $p$ to $W$; otherwise, we remove the project from consideration and repeat, until no more projects remain.

This rule aims at maximising the total utility of the voters, $\sum_{i \in N}u_i(W)$. Indeed, the Utilitarian Greedy rule is \myemph{optimal up to one project} for this objective \citep{dantzig1957discrete}, i.e., for each outcome $W$ returned by UG there exists $p \notin W$ s.t.:
\[
\sum_{i \in N}u_i(W \cup \{p\}) \geq \max_{W' \colon \cost(W') \leq b} \sum_{i \in N}u_i(W') \text{.}
\]

Note that if we use this rule with cost utilities, then the projects will be selected in descending order of their total scores (where the total score of project $p$ is $\sum_{i \in N} s_i(p)$).
This voting rule is currently used by the vast majority of cities that do participatory budgeting. We could view this choice of voting rule as a ``revealed preference'' of the city for interpreting ballots based on cost utility. If we use Utilitarian Greedy with score utilities, the rule selects projects in order of descending ``value-for-money'', which for project $p$ is
equal to $\sum_{i \in N} s_i(p)/\cost(p)$.

\paragraph{Method of Equal Shares.} This is a recent method, introduced by 
~\citet{pet-sko:laminar} and \citet{peters2021proportional}. In the first step, we divide the budget equally among the voters: for each voter $i \in N$, we set $b_i \gets \nicefrac{b}{n}$. We say that a project $p$ is \myemph{$\alpha$-affordable} for $\alpha \in \reals$ if the following equality holds:
\begin{linenomath}
\begin{equation*}
    \sum_{i\in N} \pi_i(p) = \cost(p), \text{\,where } \pi_i(p) = \min\left(b_i, \alpha u_i(p)\right).
\end{equation*}
\end{linenomath}

Here, $\pi_i(p)$ is the amount that voter $i$ needs to pay if project $p$ is selected. 
Intuitively, the condition says that the cost of project $p$ can be covered by the voters in such a way that (1) no voter exceeds their budget entitlement of $b_i$, and (2) each voter pays at most $\alpha$ per unit of utility. 
If such an $\alpha$ does not exist (which happens if and only if the total budget shares $b_i$ of the voters who assign a positive score to $p$ is lower than the cost of $p$), then the project is \myemph{not affordable}.

The method starts with $W = \emptyset$. It then repeatedly computes $\alpha$-affordability for all not-selected projects, chooses a project $p$ that is $\alpha$-affordable for minimal $\alpha$, adds $p$ to the outcome, and updates the voters' individual budgets: for each $i\in N$, $b_i \gets b_i - \pi_i(p)$. It stops when no project is affordable.

As in the case of Utilitarian Greedy, the rule can work both with
cost utilities (in which case in the first iteration it chooses an affordable project with the highest total score) and with score utilities (in which case in the first iteration it chooses
an affordable project  with the highest value-for-money). In the following iterations the rule
continues to focus on the total scores and the value-for-money, respectively, but it also takes into account how much the voters have already spent.

The Method of Equal Shares satisfies strong proportionality properties~\citep{pet-sko:laminar, peters2021proportional}, but it typically returns non-exhaustive outcomes. There are a few ways in which it can be extended to an exhaustive rule:
\begin{enumerate}[topsep=2pt, partopsep=0pt, itemsep=2pt,parsep=2pt, leftmargin=*]
    \item Completion by the Utilitarian Greedy algorithm (U): We first select an outcome $W_{\es}$ using Equal Shares; next, we select additional candidates from $C \setminus W_{\es}$ using the Utilitarian Greedy rule with the budget set to $b - \cost(W_{\es})$. We return $W_{\es}$ together with the additional candidates.
    \item Completion by tweaking voters' utilities (Eps): It is known that the outcome of Equal Shares is exhaustive if every voter's utility for every project is strictly positive~\citep{peters2021proportional}. To use this, for each $i\in N$ and each $p\in P$ with $s_i(p) = 0$, we override the voter's ballot by setting $s_i(p) \gets \varepsilon$ for some small $\varepsilon > 0$. Then, we run Equal Shares for this tweaked election.
    \item Completion by increasing the initial endowments (Add1): Observe that Equal Shares can be run with the initial voter endowments $b_i$ set to a different value than $\nicefrac{b}{n}$. We start with the endowments set to $\nicefrac{b}{n}$. If the outcome is not exhaustive, we increase the initial endowment $b_i$ by one unit, and run Equal Shares from scratch. We repeat this procedure until the outcome is exhaustive or until the moment when the next increase of  endowments would result in exceeding the original budget $b$. 

    Add1 does not necessarily return an exhaustive outcome, but the amount of unspent funds is typically small (as we will see). To get an exhaustive rule, we can combine Add1 with the Utilitarian Greedy completion; we call it Add1U.  
\end{enumerate}

\vspace{-2mm}
\section{Participatory Budgeting Library (\pabulib{})}

\pabulib{} is an open library of participatory budgeting instances that we collected in
this project (but we invite and encourage interested researchers to submit their data). 
In \Cref{sec:pb_format} we define the \emph{.pb} format, which we recommend for representing PB instances, and which is used in our library.

The aim of \pabulib{} is to gather  participatory budgeting data from as many cities and as many countries as possible, but
currently most of the instances come from several large cities in Poland (in particular, from Warsaw, with a population of 1.7 million people; from Krakow, Wroclaw, and Gdansk, with  populations between 500 000 and 1 million; and from Czestochowa, Zabrze, and Katowice, with populations between 150 000 and 300 000). 
As these cities use different voting formats, the library accepts several different types of ballots. For example, in Warsaw every voter is asked to approve up to 15 local and up to 10 citywide projects; in Krakow each voter must assign three different scores (3 points, 2 points, and 1 point) to three different projects; and in Czestochowa each voter distributes a total score of 10 points between the projects. In Wroclaw and Zabrze, a voter can only vote for a single local and a single citywide project. 

\begin{figure}[t!]
    \centering
    \begin{subfigure}[t]{0.24\textwidth}
        \includegraphics[width=0.9\textwidth]{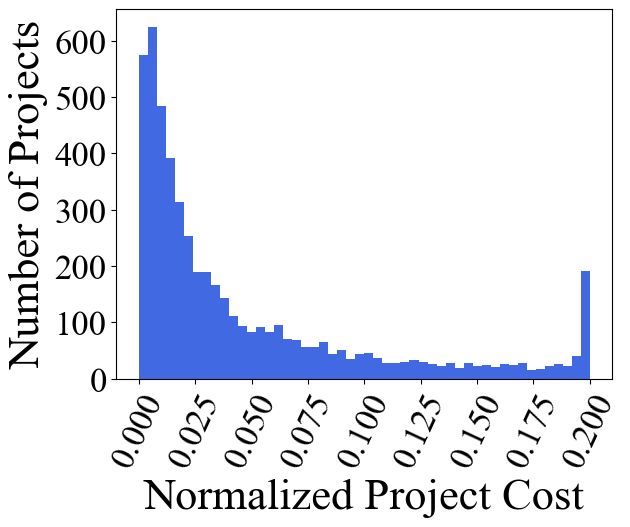}
        \caption{Warsaw 2020-2023}
    \end{subfigure}%
    \begin{subfigure}[t]{0.24\textwidth}
        \includegraphics[width=0.9\textwidth]{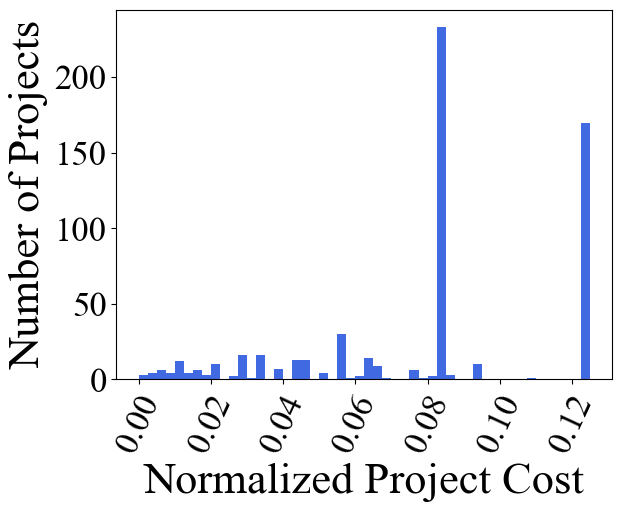}
        \caption{Wroclaw 2019-2021}
    \end{subfigure}

    \begin{subfigure}[b]{0.24\textwidth}
        \includegraphics[width=0.9\textwidth]{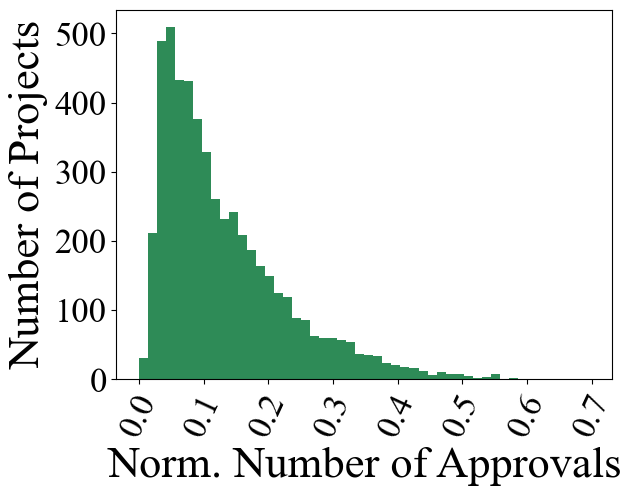}
        \caption{Warsaw 2020-2023}
    \end{subfigure}%
    \begin{subfigure}[b]{0.24\textwidth}
        \includegraphics[width=0.9\textwidth]{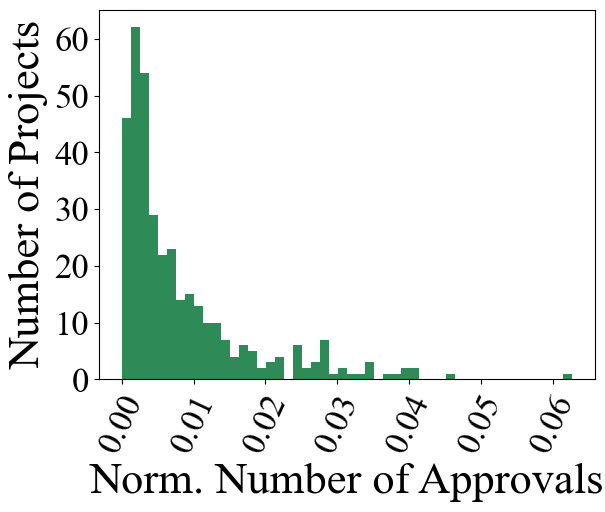}
        \caption{Wroclaw 2019-2021}
    \end{subfigure}

    \caption{\label{fig:project_cost_and_support} The distribution of costs and the total number of approvals per project for two sample cities, aggregated over different years. For each PB instance, we normalize the costs by dividing them by the total budget, and the number of approvals by dividing it by the largest number of approvals obtained by any project.}
\end{figure}


In~\Cref{fig:project_cost_and_support}, we illustrate the distribution of project costs and the total numbers of approvals per project for two sample cities, Warsaw and Wroclaw. (The distributions for other cities and the correlation between the projects costs and their support are depicted in~\Cref{fig:project_cost_and_support_apdx} in the appendix).
We observe a sharp difference between the cities, depending on the type of ballots they use. The data suggest that forcing the voters to vote only for a single project typically discourages submitting smaller and cheaper projects. We consider this effect undesirable and so this is an argument against such voting formats.



\vspace{-2mm}
\section{Basic Metrics of Fairness and Efficiency}

In this section we consider several basic metrics for comparing outcomes returned by different voting rules. 

The \myemph{average utility}, which for outcome $W$ is defined as $\nicefrac{1}{n}\sum_{i \in N} u_i(W)$, is perhaps the most natural metric of efficiency. We are also interested in the degree to which a given voting rule respects the diversity of voters' opinions, which is less straightforward to measure. One possible way would be to apply known measures of statistical dispersion (such as the Gini index) to the vector of voters' utilities: a smaller value of the dispersion would suggest that the utility is more evenly distributed, hence the opinions of a greater part of the population are taken into account. However, this approach is problematic in the context of cost utilities. Indeed, some residents choose to vote only for local, relatively cheap projects, while others support mainly large and expensive initiatives. Thus, it is expected that the cost utilities of different voters can vary substantially.
When performing the comparative analysis of two rules, $R_1$ and $R_2$, we find it more informative to compute the \myemph{dominance margin} of $R_1$ over $R_2$, i.e., the fraction of voters who enjoy a strictly higher utility from the outcome of $R_1$ than from the one of $R_2$. A related metric is the \myemph{improvement margin} of $R_1$ over $R_2$ which is the dominance margin of $R_1$ over $R_2$ minus the dominance margin of $R_2$ over $R_1$. We also measure the \myemph{exclusion ratio}---the fraction of voters who support none of the selected projects.

Next, we consider a metric recently introduced by \citet{lackner2021fairness} which, informally speaking, measures the amount of spending that different voters had an influence on \citep[][]{mal-rey-end-lac:effort-based-pb}. The measure assumes that the supporters of a selected project contributed to the decision on spending money on this project proportionally to the
score that they assigned to it. Consequently, given an outcome $W$, voter $i$'s share is:
\[
\mathrm{share}_i(W) = \sum_{p \in W} \frac{s_i(p)}{\sum_{j\in N} s_j(p)}\cdot \cost(p) \text{.}
\]
Note that the shares of the voters sum up to the total cost of the selected projects. In an ideally fair solution we would like all these shares to be equal. This suggests the following metric, which we call \myemph{power inequality}:
\begin{linenomath}
\begin{align*}
\textstyle\nicefrac{1}{n}\cdot\sum_{i\in N} |\mathrm{share}_i(W) - \nicefrac{b}{n}|\cdot  \nicefrac{n}{b} \text{.}
\end{align*}
\end{linenomath}

\begin{figure}[t!]
\includegraphics[width=\linewidth]{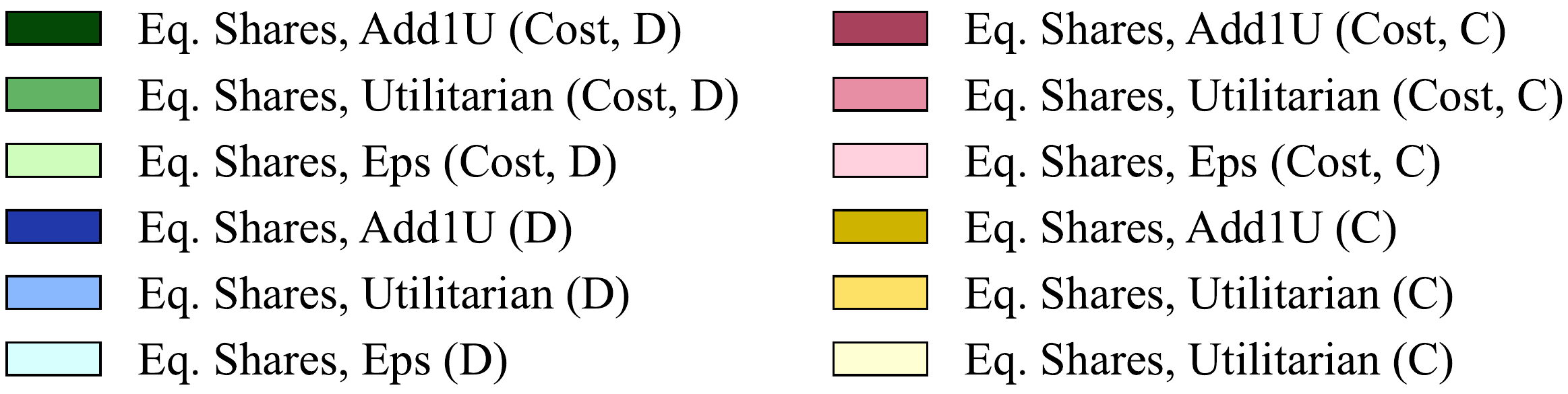}
  \includegraphics[width=\linewidth]{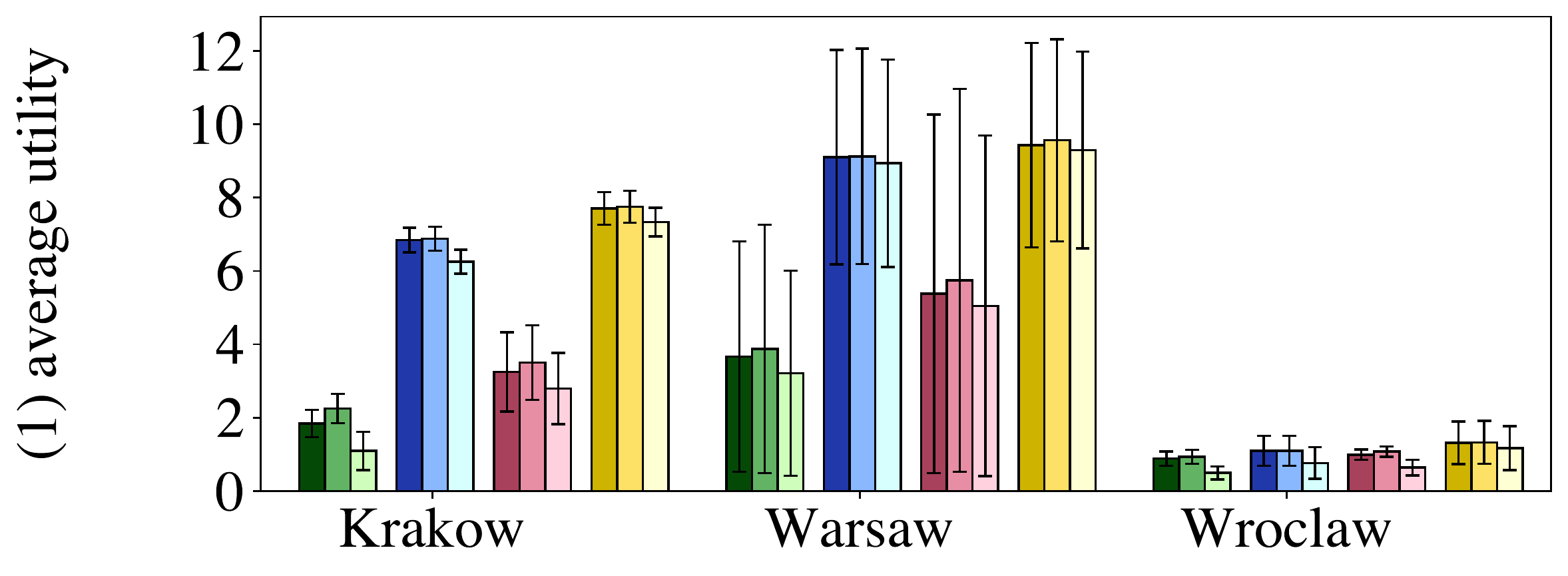}
  \includegraphics[width=\linewidth]{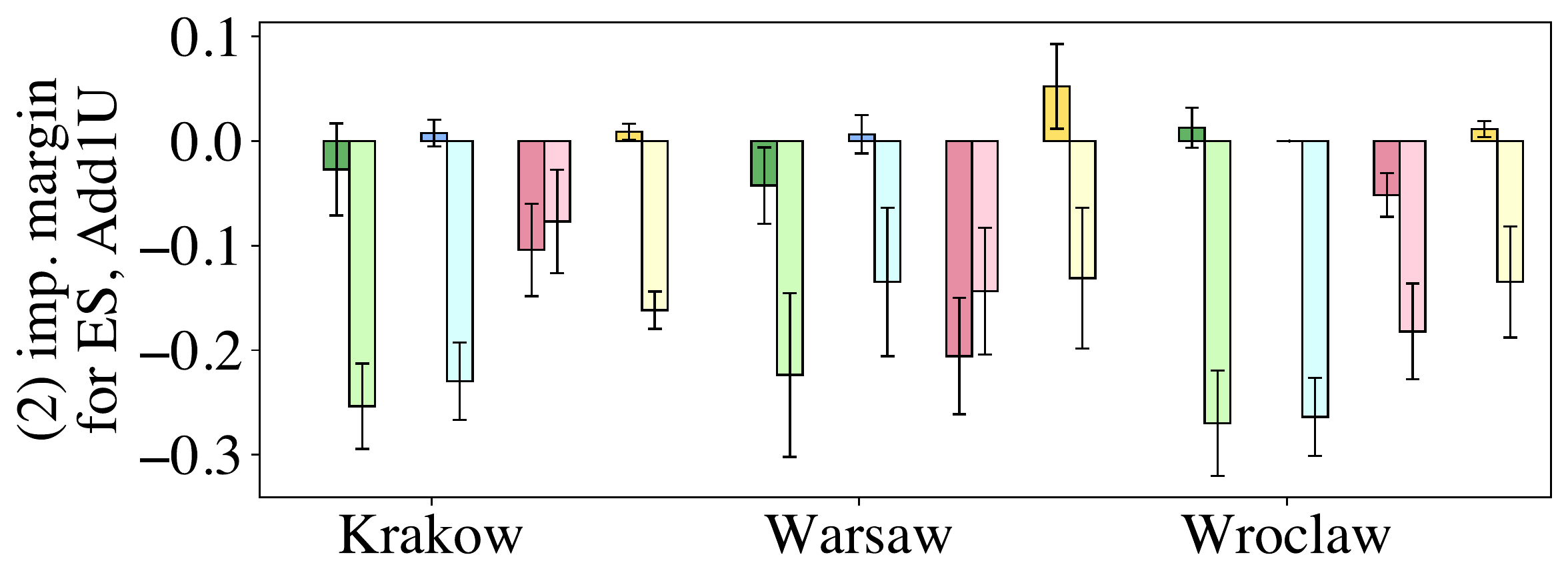}
  \includegraphics[width=\linewidth]{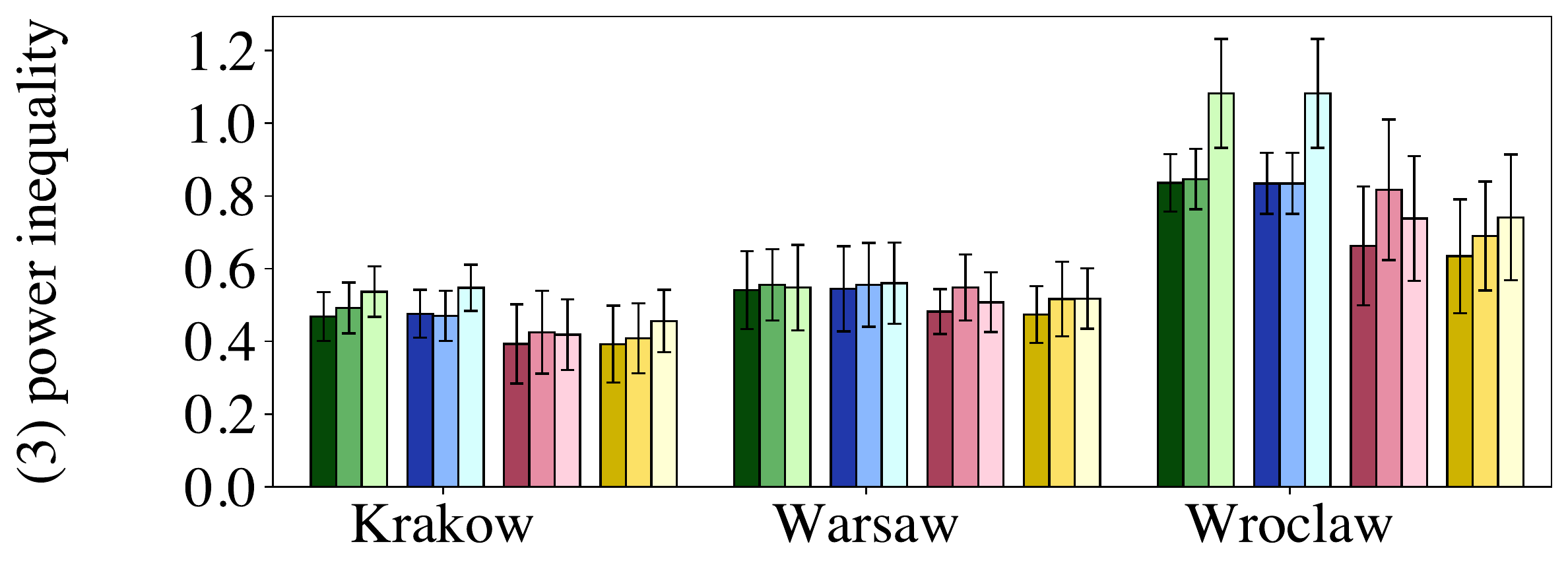}
  \caption{Comparison of different completion rules for the Method of Equal Shares. The label ``Cost'' means that we are referring to the cost-utility variant of the method; otherwise we are referring to its score-utility variant. The symbols ``D'' and ``C'' stand for the districtwise and citywide schemes, respectively. We compare (1) voters' average utility (for the score-utility variants of the methods we give the average score utility; for the cost-utility variants we give the average cost utility in millions), (2) the improvement margin over Equal Shares with the Add1U completion (the improvement margin is with respect to cost utilities for the cost-utility variants of the methods and with respect to score utilities for the remaining variants), and (3) the power inequality.}\label{fig:basic_pops_completions}
\end{figure}

We used the mentioned metrics to compare the outcomes of various election rules. We took data from PB elections carried out in seven major Polish cities: Czestochowa (2020), Gdansk (2020), Katowice (2020--2021), Krakow (2018--2022), Warsaw (2017--2023), Wroclaw (2015--2021), and Zabrze (2020--2021). In the plots we show only cities for which we have data for at least 3 years (Warsaw, Krakow, and Wroclaw). The analysis of the remaining cities led to similar conclusions.

\begin{figure*}[t!] 
  \includegraphics[width=1\linewidth]{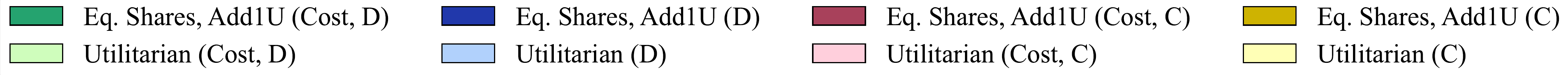}
  \includegraphics[width=0.5\linewidth]{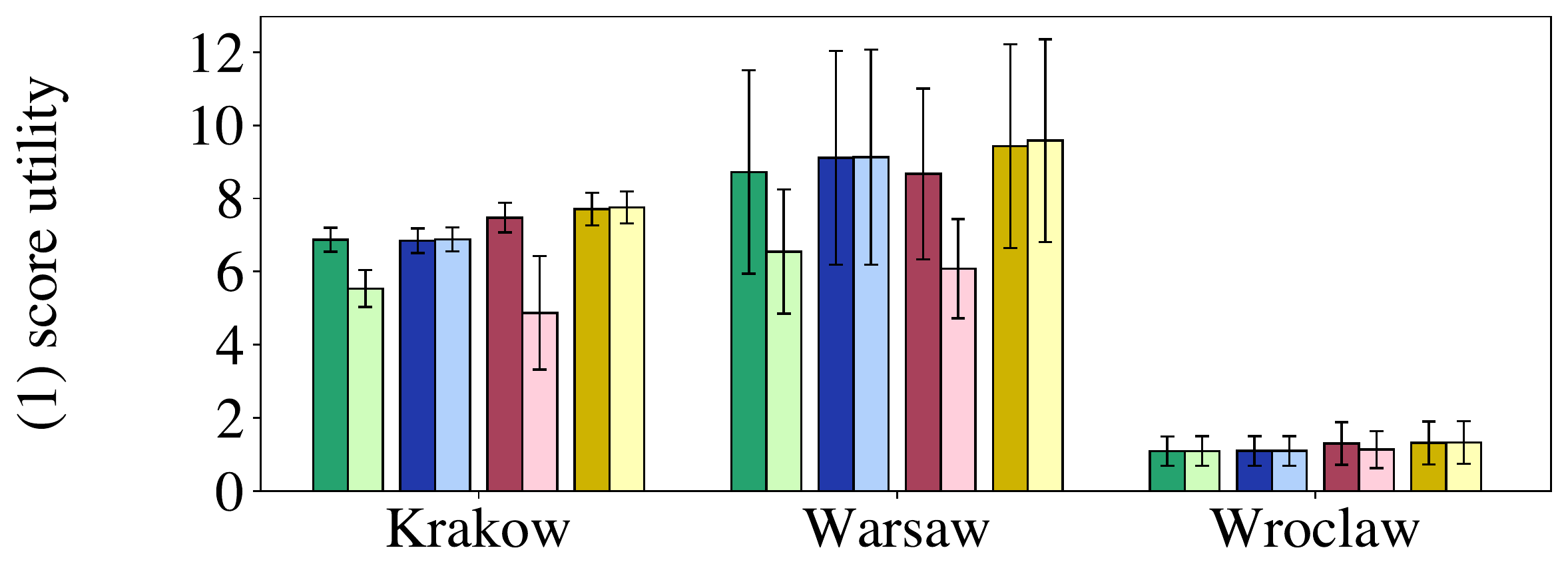}
  \includegraphics[width=0.5\linewidth]{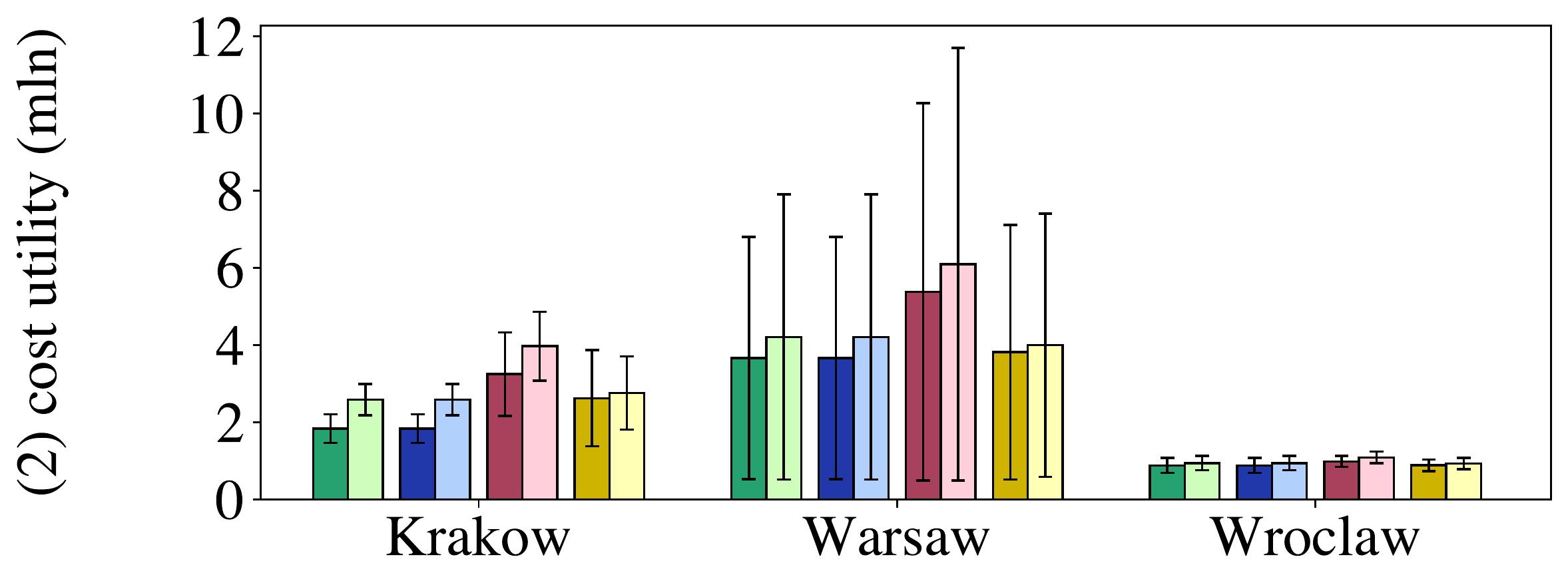}
  \includegraphics[width=0.5\linewidth]{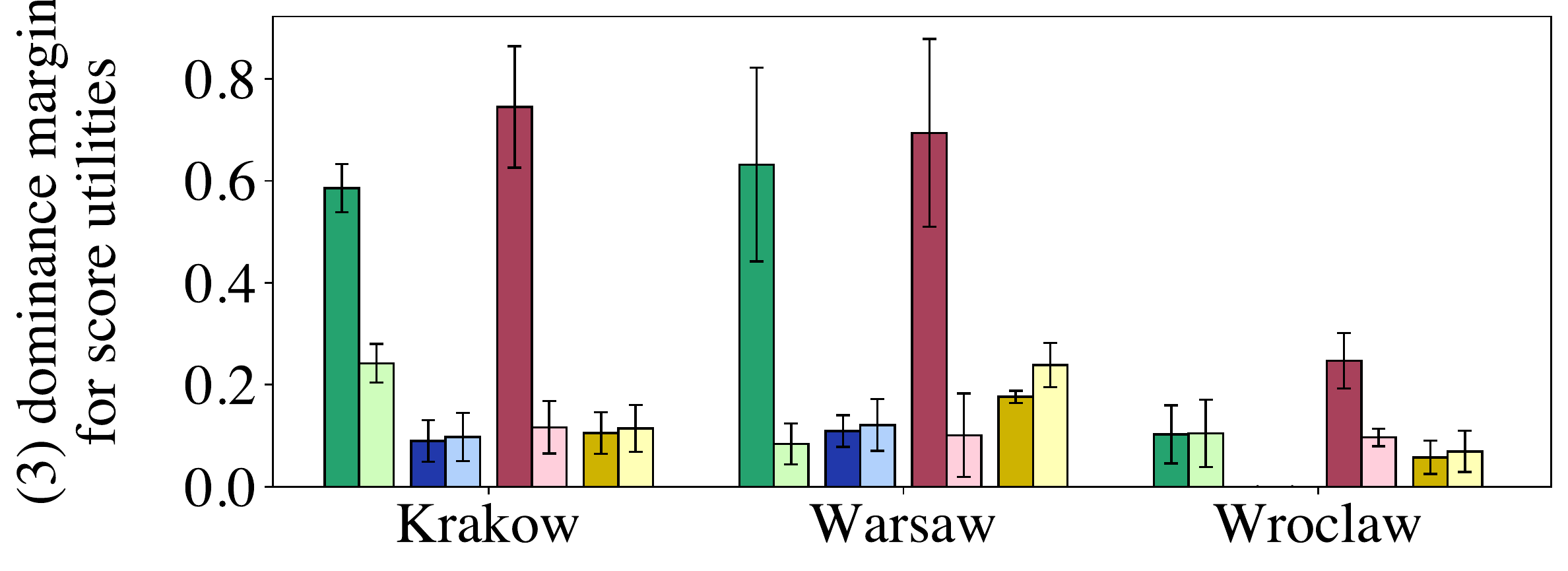}
  \includegraphics[width=0.5\linewidth]{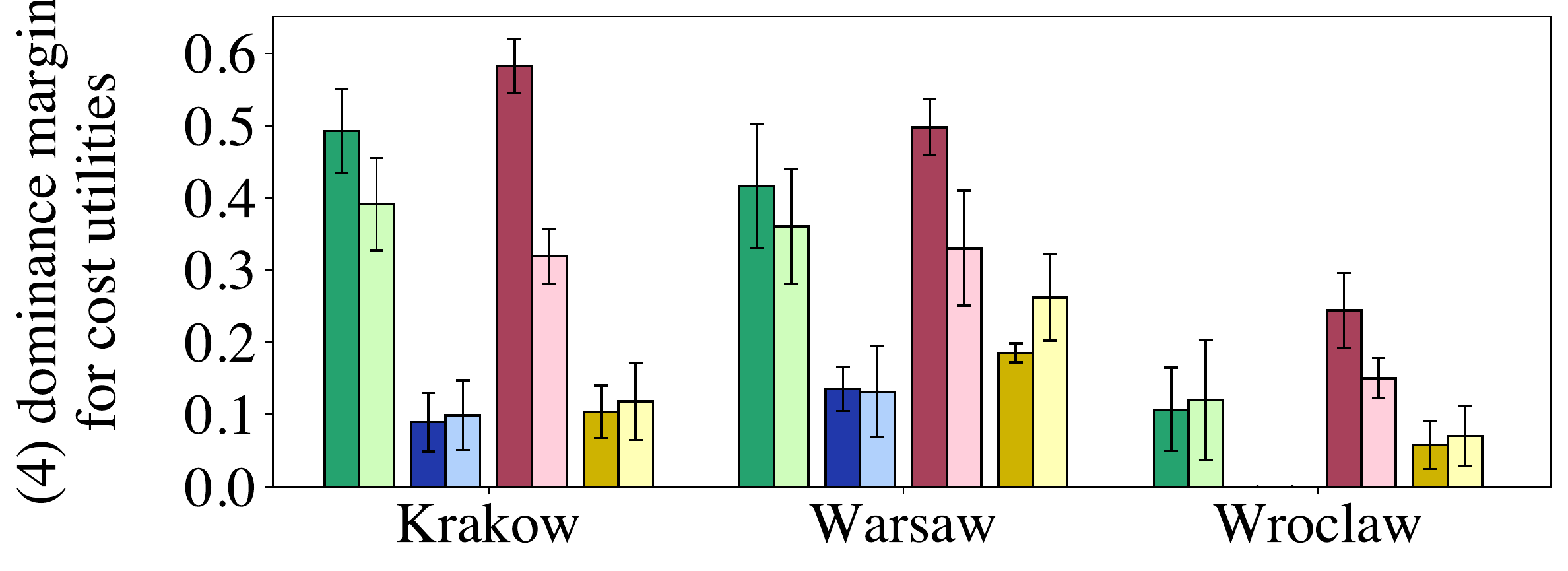}
  \includegraphics[width=0.5\linewidth]{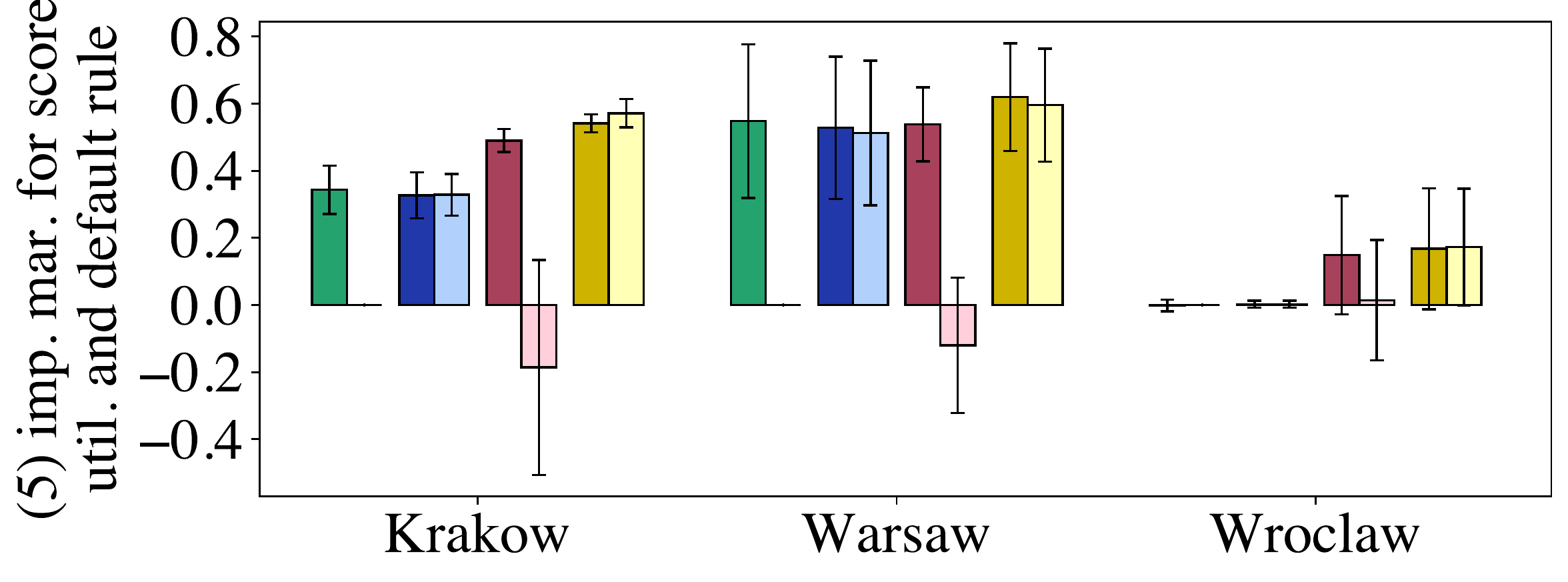}
  \includegraphics[width=0.5\linewidth]{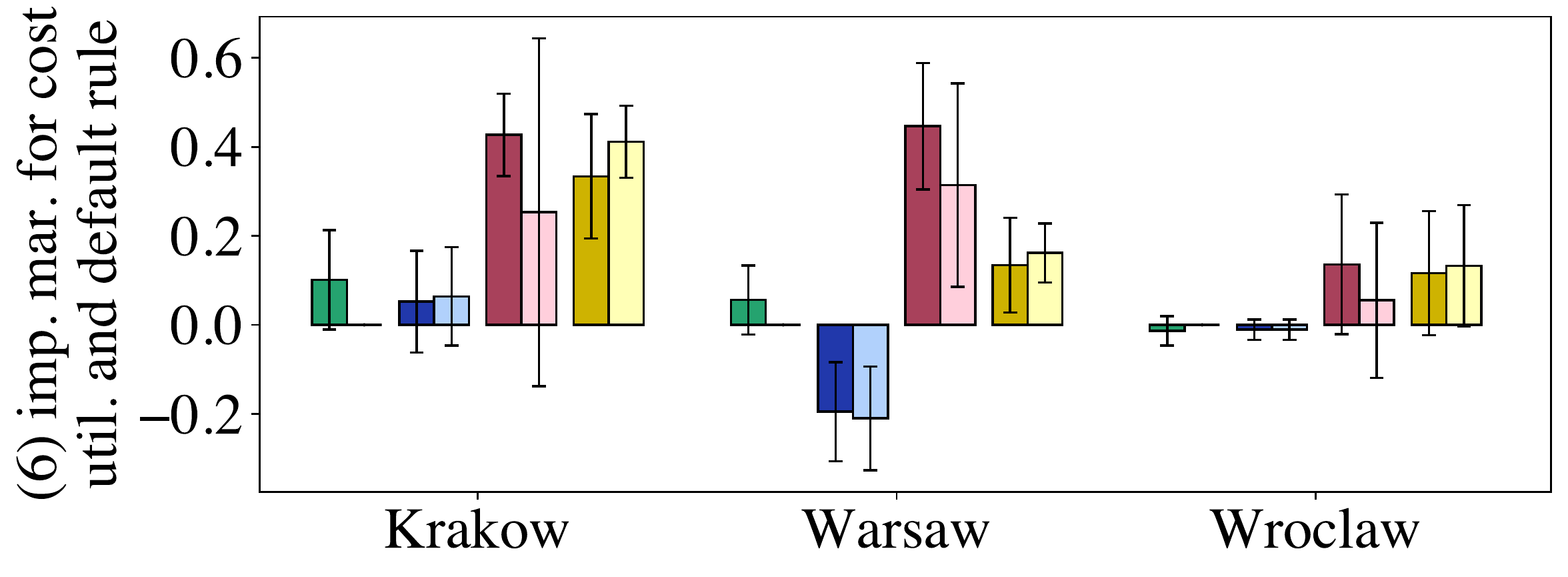}
  \includegraphics[width=0.5\linewidth]{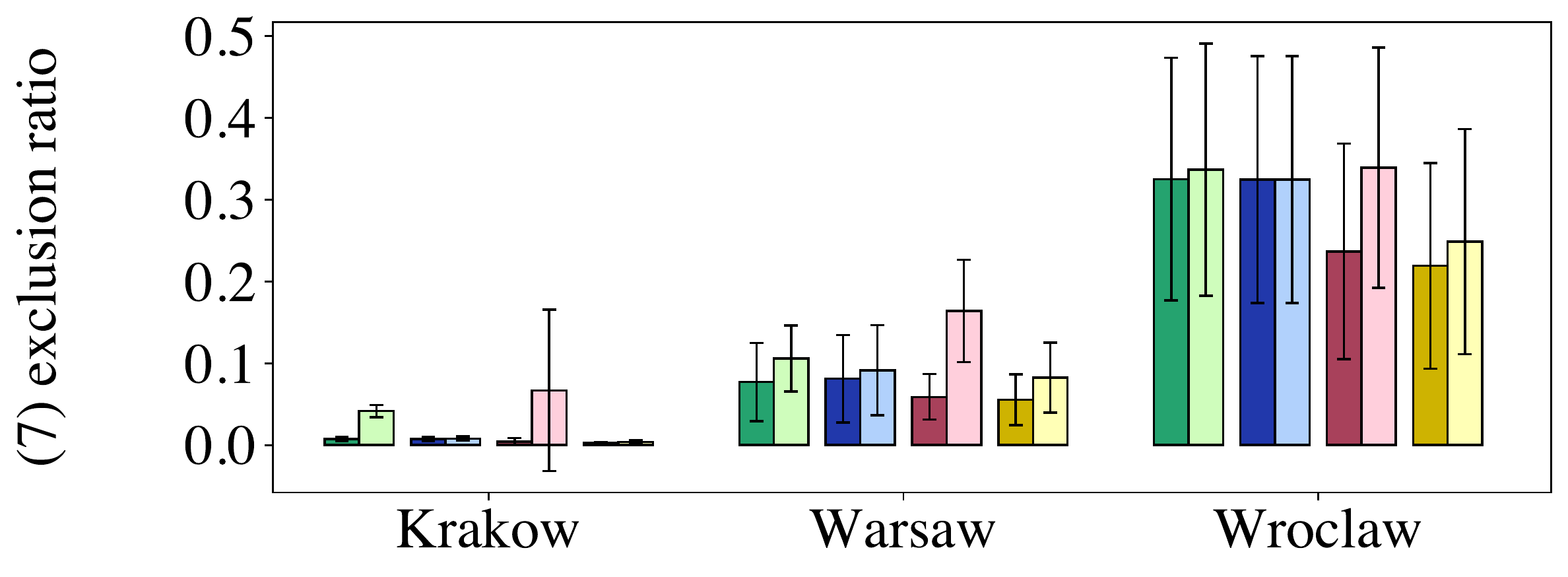}
  \includegraphics[width=0.5\linewidth]{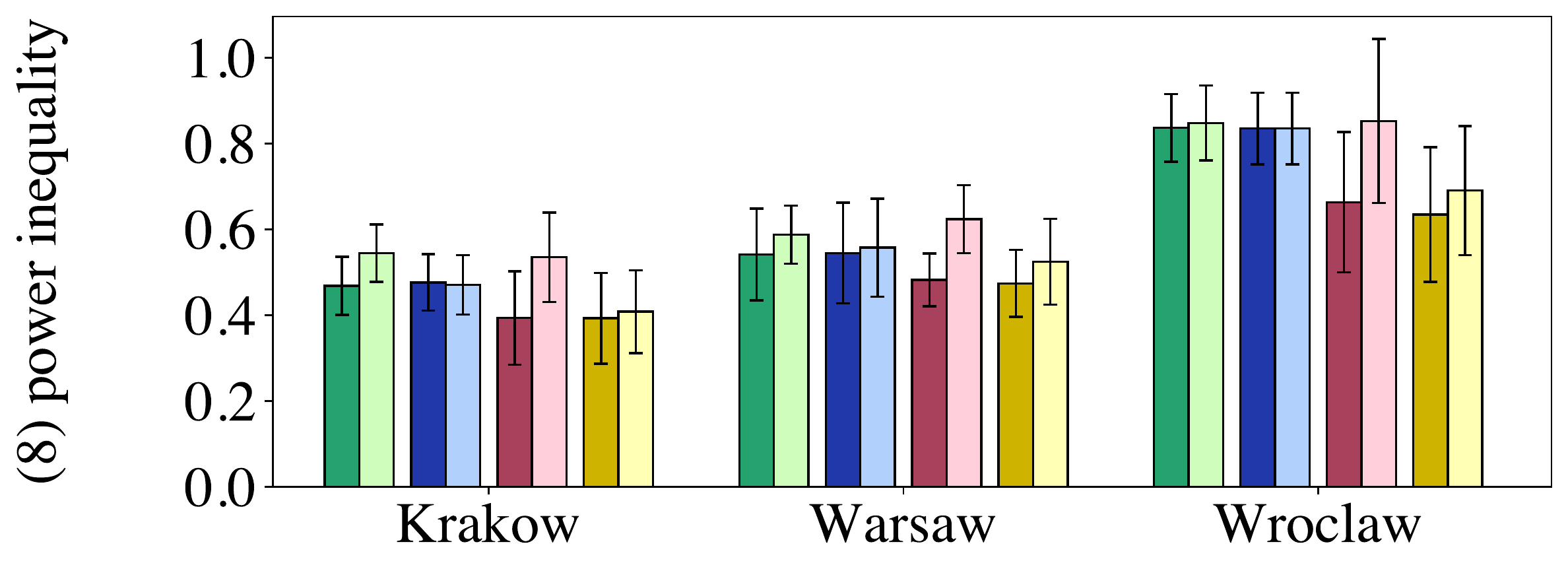}
  \caption{Comparison of Equal Shares and Utilitarian Greedy (in different variants), with respect to (1) average score utility, (2) average cost utility, (3) dominance margin with respect to score utility, (4) dominance margin with respect to cost utility, (5) improvement margin over the currently used rule (Utilitarian, Cost, D) with respect to score utilities, (6) improvement margin over the currently used rule with respect to cost utilities (7) exclusion ratio, and (8) power inequality. The label ``Cost'' means that we are referring to the cost-utility variant of the method; otherwise we are referring to its score-utility variant. The symbols ``D'' and ``C'' stand for districtwise and citywide schemes, respectively.}\label{fig:basic_pops_es_vs_ug}
\end{figure*}

To deal with city districts, for each city and each year we tested the rules using two different schemes:
\begin{enumerate}[topsep=2pt, partopsep=0pt, itemsep=2pt,parsep=2pt, leftmargin=*]
\item The \myemph{citywide} (C) scheme, where we put all the projects from different districts and categories in the same pool, and we kept the original voters' ballots. Thus, for a fixed city and year we have a single election.
\item The \myemph{districtwise} (D) scheme which corresponds to how the cities currently organize their elections: a separate election is run in each city district; typically there is also one additional election involving the same set of voters but concerning citywide projects (these are projects that are potentially interesting to voters from multiple districts). The outcome for a given city and year is obtained by adding together the outcomes of these smaller elections. The metrics are computed for this combined outcome, with respect to all the voters in the city, just like in the citywide scheme.
\end{enumerate}
When it does not lead to confusion,
we will often speak of citywide and districtwise ``elections'' rather than schemes.

We present results for each city averaged over all years, with figures showing error bars corresponding to standard deviations over the years. While we consider averages, the conclusions of our analysis also hold for each year separately. The results for separate editions can be checked through our web application: \url{http://pabulib.org/pabustats}.

In our first set of simulations, we compare the three different approaches to making the Method of Equal Shares exhaustive. We observe that the completion strategy plays a critical role. For example, in citywide and districtwise elections, Equal Shares without completion uses, on average, only 32\% and 50\% of the available funds, respectively. The Add1 completion uses on average 98\% and 88\%, and Add1U (an exhaustive variant) uses 99.9\% and 95\% of the funds, respectively. 

In 
\Cref{fig:basic_pops_completions}, we compare our metrics for the three  completion strategies, the
two types of utilities, and citywide and districtwise elections (additional plots are provided in \Cref{fig:extended_pops_completions} in the appendix). 
This figure is quite involved, so let us provide some guidance on how to read it. First, we have a separate plot
for each of the metrics, i.e., for (1) average utility, (2) the improvement margin over Equal Shares with the Add1U method, and (3)~power inequality. Within each of the plots,
different shades of each color correspond to different completion types (darkest for Add1U, middle for U, and lightest  for Eps). The green and blue shades correspond to the districtwise elections, and red and yellow shades to citywide elections. Cost utilities correspond to
green and red, whereas score utilities correspond to blue and yellow.
We conclude the following (the exclusion ratios are in the appendix):
\begin{enumerate}[topsep=2pt, partopsep=0pt, itemsep=2pt,parsep=2pt, leftmargin=*]
\item The completion by tweaking voters’ utilities (Eps, lightest shade for each color) gives the worst results in terms of the average utility, power inequality, exclusion ratio, and improvement over Add1U.
\item The utilitarian completion (U, middle shades) gives a bit higher average utility than Add1U (darkest shades), but also a worse power inequality and exclusion ratio.
\item For score utilities (blue and yellow), the utilitarian completion (U, middle shade) and Add1U (darkest shade) are comparably good. For cost utilities (green and red), the Add1U strategy (darkest shade) has a large advantage, especially for improvement margins (the other rules have negative values in almost all settings).
\end{enumerate}
Based on these observations, we suggest  using the Add1U completion as the default option.

In our second set of simulations we compare the Add1U variant of the Method of Equal Shares with the Utilitarian Greedy rule. The results are depicted in \Cref{fig:basic_pops_es_vs_ug} (see also \Cref{fig:extended_pops_es_vs_ug} in the appendix). 
In this figure, each scenario corresponds to a color. The darker shades represent Equal Shares and
the lighter ones represent Utilitarian Greedy. Our findings can be summarised as follows:
\begin{enumerate}[topsep=2pt, partopsep=0pt, itemsep=2pt,parsep=2pt, leftmargin=*]
\item For rules based on score utilities (blue and yellow), the results of the Method of Equal Shares and of Utilitarian Greedy are comparable. Equal Shares selects outcomes with slightly lower exclusion ratio as well as lower power inequality, at the cost of lower average utility. However, these differences are relatively small.  
\item For rules based on cost utilities (green and red), we see a significant difference between the two rules. Unsurprisingly, the outcomes of Utilitarian Greedy have better average cost utility, but the difference is relatively small (e.g., it is the largest in Warsaw for citywide elections, 13\%). For all other metrics, Equal Shares outperforms Utilitarian Greedy by a large margin. For example, for citywide elections in Warsaw, the average score utility of the outcomes of Equal Shares is 43\% higher, and using Equal Shares would result in a drop of the exclusion ratio from 16\% to 6\%. The improvement margin over UG in terms of score and cost utility is on average, respectively, 59\% and 17\%.
\item We observe a significant difference between the citywide (red and yellow) and districtwise (blue and green) schemes. Global elections (that do not divide projects \emph{a priori} into districts) result in a much higher average utility and much lower exclusion ratio, for Equal Shares.\footnote{The large difference between districtwise and citywide elections arises because in some districts no popular projects are submitted, and their residents would prefer to fund citywide projects instead. 
As an example, in 2021, Warsaw had 2 projects about the same street: ($A$) new plants for Modlinska street (citywide project), and ($B$) new pavement for Modlinska street (district project in Bialoleka).  Project $A$ received six times as many votes as $B$ (12 463 vs 1 932). Even among the voters of Bialoleka, $A$ was twice as popular (4 365 vs 1 932). Project $A$ was cheaper (435k vs 630k PLN). Even though $A$ was more popular and cheaper than $B$, it was not selected (being a citywide project) while $B$ was selected.} 
\end{enumerate}

Finally, it is interesting to see how fairly the budgets were distributed among districts when using the citywide scheme. Let $\mathcal{D}=\{D_1, D_2, \ldots, D_t\}$ be the set of districts, which is formally a partition of $N$. Ideally, the voters from a district $D \in \mathcal{D}$ should get a share of the budget that is proportional to the size of $D$.\footnote{This assumes that the budget should be divided proportionally to the number of voters and not to the number of residents of a district. If turnout varies between districts, the difference matters. Being proportional to the number of voters promotes participation, incentivizes districts to encourage their residents to vote, and follows the ``one person, one vote'' principle. If the city prefers being proportional to the number of residents, the citywide scheme can be adapted by giving voters from districts with lower turnout a larger initial endowment.} The \myemph{dispersion of the budget allocation} is:
\[
\frac{1}{|\mathcal{D}|} \cdot \sum_{D \in \mathcal{D}} \frac{\big|\sum_{i \in D}\mathrm{share}_i(W) -  \nicefrac{|D|}{n}\cdot b\big|}{\nicefrac{|D|}{n}\cdot b}\text{.}
\]
This metric captures the average relative difference between how much money the district got and how much we would expect it to get. \Cref{tab:district-strength} shows average dispersion values, which are lower for Equal Shares than for Utilitarian Greedy. 

\newcommand{\numberbarDisperson}[1]{\tikz{
	\fill[red!80!black!16] (0,0) rectangle (#1*35mm,7pt);
	\node[inner sep=0pt, anchor=south west] at (0,0) {#1};}
}

\newcommand{\numberbarDispersontrunc}[1]{\tikz{
		\fill[red!80!black!16] (0,0) rectangle (17.5mm,7pt);
		\node[inner sep=0pt, anchor=south west] at (0,0) {#1};
		\draw[white, thick] (16.5mm, 7pt) -- (15.5mm, 4.5pt) -- (16mm, 3pt) -- (15mm, 0pt);}
}

\begin{table}[t]
    \centering
\begin{tabular}{p{20mm}|p{16mm}|p{16mm}|p{16mm}}
    City & Add1U, C & Util. G, D & Util. G, C \\
\midrule
    Czestochowa & \numberbarDisperson{0.23} & \numberbarDisperson{0.28} & \numberbarDisperson{0.39} \\
    Gdansk      & \numberbarDisperson{0.27} & \numberbarDisperson{0.33} & \numberbarDisperson{0.46} \\
    Katowice    & \numberbarDisperson{0.19} & \numberbarDisperson{0.26} & \numberbarDisperson{0.51} \\
    Krakow      & \numberbarDisperson{0.08} & \numberbarDisperson{0.24} & \numberbarDisperson{0.23} \\
    Warsaw      & \numberbarDisperson{0.20} & \numberbarDisperson{0.41} & \numberbarDisperson{0.41} \\
    Wroclaw     & \numberbarDisperson{0.15} & \numberbarDisperson{0.26} & \numberbarDisperson{0.22} \\[-0.5pt]
    Zabrze      & \numberbarDisperson{0.38} & \numberbarDispersontrunc{1.24} & \numberbarDisperson{0.41}
\end{tabular}
    \caption{Average dispersion of the budget allocation. We compare cost-utility variants of Equal Shares (with the Add1U completion) and Utilitarian Greedy (on the districtwise and the citywide schemes).}
    \label{tab:district-strength}
\end{table}

\section{Robustness to Changing the Type of Ballots}
Recently, \citet{ben-fai-gal:pb-real-experiments} performed lab experiments about different input formats for PB elections. One of their findings was that the Method of Equal Shares is robust to changing the type of ballots. In this section we reinforce their conclusions by analyzing data from real PB elections.

For each election where voters used cardinal ballots, we construct a corresponding approval election by letting a voter approve all the projects to which she assigned a positive score. Then, we compare the outcomes of different rules for these two elections. Let $W_{\sc}$ and $W_{\mathrm{appr}}$ be the outcomes of a given voting rule for the original and the approval elections, respectively. We define the \myemph{robustness ratio} as $\cost(W_{\mathrm{appr}} \cap W_{\sc}) / \cost(W_{\sc})$. \Cref{tab:robustness} summarises the results of our analysis. We can see that the outcomes of Equal Shares change much less after switching to approval compared to Utilitarian Greedy. For users of Equal Shares, this provides an argument in favor of approval ballots, which is the voting format that \citet{ben-fai-gal:pb-real-experiments} find to be the easiest to use.

\newcommand{\numberbarRobustness}[1]{\tikz{
	\fill[blue!17] (0,0) rectangle (#1*19mm,7pt);
	\node[inner sep=0pt, anchor=south west] at (0,0) {#1};}
}
\begin{table}[t]
    \centering
\begin{tabular}{p{20mm}|p{16mm}|p{16mm}|p{16mm}}
    City & Add1U, C & Util. G, D & Util. G, C \\
\midrule
    Czestochowa & \numberbarRobustness{0.80} & \numberbarRobustness{0.35} & \numberbarRobustness{0.39} \\
    Gdansk      & \numberbarRobustness{0.87} & \numberbarRobustness{0.26} & \numberbarRobustness{0.39} \\
    Katowice    & \numberbarRobustness{0.83} & \numberbarRobustness{0.56} & \numberbarRobustness{0.42} \\
    Krakow      & \numberbarRobustness{0.78} & \numberbarRobustness{0.52} & \numberbarRobustness{0.41} 
\end{tabular}
    \caption{Robustness ratio for different voting rules. We compare cost-utility variants of Equal Shares (with the Add1U completion) and Utilitarian Greedy (on the districtwise and the citywide schemes).}
    \label{tab:robustness}
\end{table}

\section{Budget Distribution among Categories}    

Cities often organize projects by topics (such as public space, environment, education) to make browsing the list of projects easier. Warsaw, for example, categorizes projects using 10 different tags (where projects can get multiple tags). This allows us to ask whether voter preferences across categories are well-reflected by the spending of the rules.

\begin{figure}
    \centering
    \definecolor{piece1}{RGB}{32, 48, 128}
\definecolor{piece2}{RGB}{51, 72, 181}
\definecolor{piece3}{RGB}{60, 90, 240}
\definecolor{piece4}{RGB}{171, 24, 24}
\begin{tikzpicture}
    [barlabel/.style={anchor=south west, font=\small, inner sep=0, scale=0.97},
    taglabel/.style={anchor=south west, font=\scriptsize, inner sep=0, scale=0.87},
    piecelabel/.style={anchor=south west, font=\footnotesize, inner xsep=1pt, inner ysep=4.5pt, text=white, scale=0.68}]

    \node [barlabel] at (0, 1.1) {Vote share\phantom{g}};
    \node [barlabel] at (0, 0.5) {Equal Shares};
    \node [barlabel] at (0, 0) {Utilitarian Greedy};
    
    \draw [black!15] (0,1) -- (8.5,1);
    
    \node [taglabel] at (2.45, 1.57) {Public space};
    \node [taglabel] at (3.7, 1.57) {Sport};
    \node [taglabel] at (4.63, 1.57) {Transit\phantom{p}};
    \node [taglabel] at (7.62, 1.57) {Education\phantom{p}};
    
    
    \foreach \x/\piececolor/\label [remember=\x as \lastx (initially 0)] in {
            0.205/piece1/20\%, 
            0.363/piece2/16\%, 
            0.411/piece3/5\%,
            0.497/black!10/,
            0.555/black!10/,
            0.573/black!10/,
            0.585/black!10/,
            0.669/black!10/,
            1/piece4/
        }
        {
            \fill [fill=\piececolor] (2.45 + \lastx*6.05,1.1) rectangle (2.45 + \x*6.05, 1.5);
            \draw [black!40, very thin] (2.45 + \x*6.05, 1.1) -- (2.45 + \x*6.05, 1.5);
            \node [piecelabel] at (2.45 + \lastx*6.05,1.1) {\label};
        }
    \foreach \x/\piececolor/\label [remember=\x as \lastx (initially 0)] in {
            0.237/piece1/24\%, 
            0.422/piece2/19\%, 
            0.528/piece3/11\%,
            0.611/black!10/,
            0.673/black!10/,
            0.685/black!10/,
            0.691/black!10/,
            0.739/black!10/,
            1/piece4/
        }
        {
            \fill [fill=\piececolor] (2.45 + \lastx*6.05,0.5) rectangle (2.45 + \x*6.05, 0.9);
            \draw [black!40, very thin] (2.45 + \x*6.05, 0.5) -- (2.45 + \x*6.05, 0.9);
            \node [piecelabel] at (2.45 + \lastx*6.05,0.5) {\label};
        }
    \foreach \x/\piececolor/\label [remember=\x as \lastx (initially 0)] in {
            0.300/piece1/30\%, 
            0.564/piece2/26\%, 
            0.717/piece3/15\%,
            0.821/black!10/,
            0.910/black!10/,
            0.920/black!10/,
            0.921/black!10/,
            0.946/black!10/,
            1/piece4/
        }
        {
            \fill [fill=\piececolor] (2.45 + \lastx*6.05,0) rectangle (2.45 + \x*6.05, 0.4);
            \draw [black!40, very thin] (2.45 + \x*6.05, 0) -- (2.45 + \x*6.05, 0.4);
            \node [piecelabel] at (2.45 + \lastx*6.05,0.0) {\label};
        }
    
    \node [piecelabel, anchor=south east] at (2.45 + 6.05,1.1) {33\%};
    \node [piecelabel, anchor=south east] at (2.45 + 6.05,0.5) {26\%};
    \node [piecelabel, anchor=south east] at (2.45 + 6.06,0) {5\%};
        
    \draw (2.45,1.1) rectangle (8.5,1.5);
    \draw (2.45,0.5) rectangle (8.5,0.9);
    \draw (2.45,0) rectangle (8.5,0.4);
    \end{tikzpicture}
    \caption{The vote share and the spending share of different tags in district Bielany, Warsaw 2020. The picture is similar for 2021--23.}
    \label{fig:bielany-categories}
\end{figure}
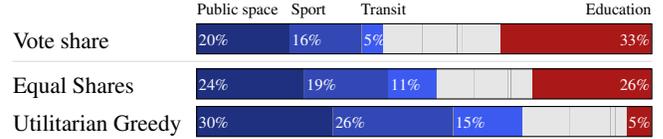

\begin{figure*}[t!h]
    \centering
    \includegraphics[width=16cm]{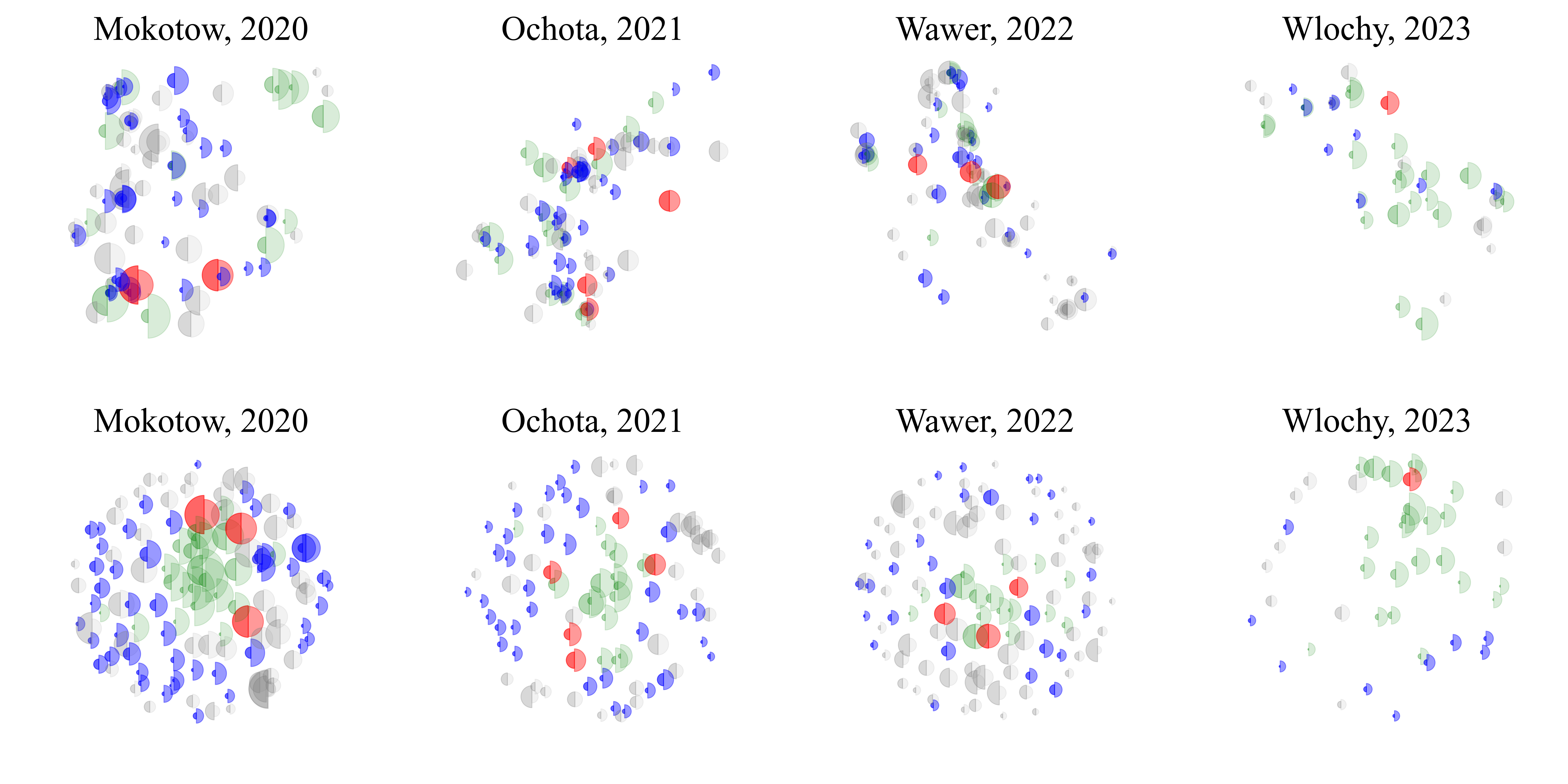}
    \caption{\label{fig:maps_of_elections} Visualisation of projects in PB elections using GPS data (upper row) and using the Jaccard distance (lower row). Each 
    project is represented by two glued-together half-discs. The size of the left half    
    is proportional to the project's cost, whereas the size of the right half is proportional to the total number of votes the project received. The figures compare the outcomes of the cost-utility variant of Equal Shares with Add1U completion, with the outcomes of the Utilitarian Greedy rule.
    Specifically, gray projects were not selected by either of the rules,  green projects were selected by both, blue projects were selected only by Equal Shares, and red projects were selected only by Utilitarian Greedy.}\vspace{-1mm}
\end{figure*}


\newcommand{\tags}{\mathrm{tags}}
We focus on Warsaw district elections (2020--23), which use approval voting. Denote by $A_i$ the set of projects approved by $i \in N$. For each project $p$, denote by $\tags(p) \subseteq \{\text{public space}, \dots, \text{education}\}$ the tags assigned to $p$. For each tag $t$, we can then compute its \myemph{vote share}:
\[
\frac{1}{n}
\sum_{i\in N} \sum_{p \in A_i : t \in \tags(p)} \frac{1}{|A_i| \cdot |\tags(p)|}
\]
This intuitively counts the fraction of the votes that went to projects with tag $t$, in a way that each voter contributes the same amount to the vote share, and for projects with multiple tags, splitting their contribution equally between them. Note that the vote shares of all the tags sum to 1. For an outcome~$W$, we can similarly define the spending share of the tag: $\frac{1}{\cost(W)}\sum_{p \in W : t \in \tags(p)} \cost(p)/|\tags(p)|$.

We can now compute the $\ell_2$ distance between the vector of vote shares and the spending shares of all the tags, to see how well they align. While it is not necessarily desirable for the two vectors to perfectly coincide, a large distance indicates that an outcome neglects certain categories.

We find that for 93\% of districts, Utilitarian Greedy gives outcomes with a larger distance to the vote shares than the Equal Shares outcome (cost utilities, Add1U), see \Cref{fig:division_into_categories} in the appendix. In some cases, the distance is much larger, like in the district Bielany, where in each year, Utilitarian Greedy spends much less on education projects than suggested by the vote shares. For example, in 2020, when education had a vote share of 33\%, Utilitarian Greedy spent only 5\% of the budget on these projects (Equal Shares spent 26\%), see \Cref{fig:bielany-categories}.

\section{Maps of Participatory Budgeting Elections}

We provide easy-to-interpret visualisations of the outcomes of different voting rules. For elections that were carried out in Warsaw between 2020 and 2023, most of the projects (but not all) were associated with their GPS locations. Thus, we can depict those submitted projects 
that have GPS data 
in such a way that their relative locations correspond to their physical locations in the respective districts. We present such visualisations in \Cref{fig:maps_of_elections} (additional figures are given in \Cref{sec:maps_of_elections_apdx}).

A different approach is to create a map that illustrates  voters' preferences rather than 
geographic locations of the projects. Here, for a given approval PB election  we first compute the Jaccard distances between all pairs of projects. Recall that for two projects, $p_1$ and $p_2$, their Jaccard distance is~$|N(p_1) \triangle N(p_2)|/|N(p_1) \cup N(p_2)|$, where $N(p)$ denotes the set of voters who support project $p$ (in other words, we assume that two projects are similar if similar groups of voters voted for them). Next, based on these distances, we create a two-dimensional embedding, using the Multidimensional Scaling Algorithm MDS\footnote{All the distances lie between $0$ and $1$, but most of them are relatively high, with very few being below 0.5. Thus, we normalize the distances by subtracting 0.5, i.e., $d' = max(0, d-0.5)$.} 
\citep{kruskal1964multidimensional,de2005modern}. 
Eventually, we obtain a plot where the closer two projects are, the larger is the fraction of their common supporters
(however we do mention that MDS is only a heuristic and, more importantly, a perfect embedding may not exist, so these plots provide intuitions, but should by interpreted with care).

These two types of maps for four sample elections are depicted in \Cref{fig:maps_of_elections}. Other elections are visualised in \Cref{sec:maps_of_elections_apdx}. 
We observe that Equal Shares selects more diverse and more representative sets of projects both in terms of their geographic locations and
in terms of their supporters. We further observe that Utilitarian Greedy mainly selects large and expensive projects, whereas Equal Shares selects a mixture of projects, including some large and some small ones.


\vspace{-3mm}
\section{Conclusions}
\vspace{-1mm}

Our results lead to several conclusions that should be taken into account by election designers. First, the types of ballots to be used affect the process of preparing project proposals. For example, voting for single projects discourages submitting small and medium proposals. Second, we see that the cost-utility variants of the rules select fewer but larger projects compared to their score-utility counterparts. Third, the proportional methods such as the cost-utility variants of Equal Shares result in much fairer solutions, where the different voters' opinions are better represented; the difference is substantial. Fourth, running citywide elections, without dividing the budget and the projects \emph{a priori} between districts and/or categories results in a significant improvement in terms of the average utility as well as the number of voters whose opinions are taken into account. Fifth, Equal Shares is more robust to changing the ballot types than the utilitarian rule.

\section*{Acknowledgments}
Grzegorz Pierczy\'nski and Piotr Skowron were supported by Poland's National Science Center grant no. 2019/35/B/ST6/02215.
Nimrod Talmon has been supported by the Israel Science Foundation (grant No.
630/19). Dariusz Stolicki has been supported under the Poland's National Science Center grant no. 2019/35/B/HS5/03949 and Jagiellonian's University Excellence Initiative, project QuantPol. Stanisław Szufa was supported by NCN project 2018/29/N/ST6/01303.
This project has received funding from the European Research Council
(ERC) under the European Union’s Horizon 2020 research and innovation
programme (grant agreement No 101002854).
\begin{center}
  \includegraphics[width=3cm]{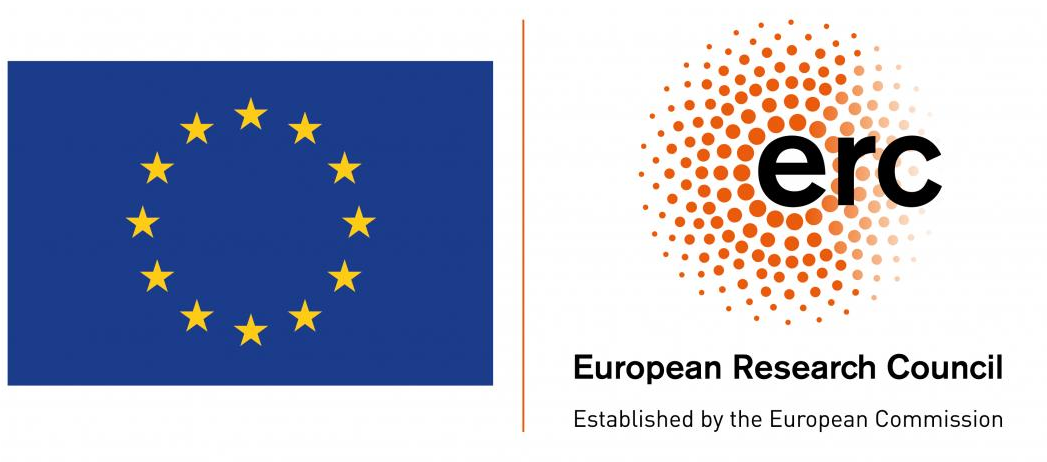}
\end{center}

\bibliographystyle{named}
\bibliography{biblio}

\clearpage
\appendix
\onecolumn
\nolinenumbers
\sectionfont{\centering}
\renewcommand\thefigure{\thesection.\arabic{figure}}

\section{\pabulib{}: Data Format}\label{sec:pb_format}
In this section we define the \emph{.pb} format, which we recommend for storing PB instances. 

The data concerning a single instance of participatory budgeting is stored in a single UTF-8 text file with the extension \texttt{.pb}.
The file should consists of three sections:
\begin{description}
    \item[META] section containing general information about the election, such like the country, the budget, and the number of votes.
    \item[PROJECTS] section specifying the costs of the projects and optionally providing additional information about the projects, such as their categories.
    \item[VOTES] section listing all votes cast in the election, optionally with additional information about the respective voters (e.g., their age, sex, etc.). We support four types of ballots: approval, ordinal, cumulative, and scoring.
\end{description}

\subsection{A Toy Example}

\begin{Verbatim}[frame=single]
META 
key; value
description; Municipal PB in Wieliczka
country; Poland
unit; Wieliczka
instance; 2020
num_projects; 5
num_votes; 10
budget; 2500
rule; greedy
vote_type; approval
min_length; 1
max_length; 3
PROJECTS
project_id; cost; category 
1; 600; culture, education 
2; 800; sport
4; 1400; culture
5; 1000; health, sport
7; 1200; education
VOTES
voter_id; age; sex; vote
1; 34; f; 1,2,4
2; 51; m; 1,2
3; 23; m; 2,4,5
4; 19; f; 5,7
5; 62; f; 1,4,7
6; 54; m; 1,7
7; 49; m; 5
8; 27; f; 4
9; 39; f; 2,4,5
10; 44; m; 4,5
\end{Verbatim}

\subsection{Detailed Description}

The fields marked with the \textbf{bold} font are obligatory.

\subsubsection{META}
    
    \begin{itemize}
        \item \bftt{key}
         \begin{itemize}
            \item \bftt{description}
            \item \bftt{country}
            \item \bftt{unit}: the name of the municipality, region, organization, etc., holding the PB process.
            \item \texttt{subunit}: the name of the sub-jurisdiction or category of the particular election.
            \begin{itemize}
                \item \textit{Example}: in Paris, a single edition of participatory budgeting consists of 21 independent elections---there is one election concerning city-wide projects and 20 local elections, one per each district. For the citywide election, the field \texttt{unit} is set to Paris, and \texttt{subunit} is undefined; for the district elections, the field \texttt{unit} is also Paris, and the \texttt{subunit} is the name of the respective district (e.g., IIIe arrondissement).
                \item \textit{Example}: before 2019, in Warsaw there were two types of local elections: district elections and neighborhood elections. For all of them, the field \texttt{unit} is set to Warsaw; the field \texttt{subunit} is the name of the district (for district elections) or the name of the neighborhood (for neighborhood elections). In order to connect neighborhoods with their districts, an optional field \texttt{district} can be used.
                \item \textit{Example}: suppose that in a given city, there is a separate election for each of $n>1$ categories (e.g., environmental projects, transportation projects, cultural projects, etc.). For each such an election the field \texttt{unit} is set to the city name; the field \texttt{subunit} is set to the name of the respective category.
            \end{itemize}
            \item \bftt{instance}: a unique identifier of the participatory budgeting edition (e.g., year, edition number, etc.). Note that the year specified in the field \texttt{instance} does not necessarily correspond to the year in which the elections were held---some organizers identify the edition by the fiscal year in which the projects are carried out.
            \item \bftt{num\_projects}
            \item \bftt{num\_votes}
            \item \bftt{budget}: the total amount of funds     
            \item \bftt{vote\_type}: the type of ballots used in the election. The library currently supports four types of ballots:
            \begin{itemize}
                \item \texttt{approval}: each vote is a vector of Boolean values, $\mathbf{v} \in \{0, 1\}^{|P|}$, where $P$ is the set of all projects,
                \item \texttt{ordinal}: each vote is a permutation of a subset $Q \subseteq P$ such that $|Q| \in [\mathtt{min\_length}, \mathtt{max\_length}]$, corresponding to a strict preference order over $Q$,
                \item \texttt{cumulative}: each vote is a vector $\mathbf{v} \in \mathbb{R}_{+}^{|P|}$ such that $\sum_{p \in P}v[p] \le \mathtt{max\_sum\_points} \in \mathbb{R}_{+}$,
                \item \texttt{scoring}: each vote is a vector $\mathbf{v} \in I^{|P|}$, where $I \subseteq \mathbb{R}$.
            \end{itemize}
            \item \bftt{rule}: the name of the rule that was used in the election. Currently we support the following rules: 
            \begin{itemize}
                \item \texttt{greedy}: this corresponds to the greedy utilitarian rule with cost utilities,
                \item other rules will be defined in the future versions.
            \end{itemize}
            \item \texttt{date\_begin}: the date when the process of collecting ballots started.
            \item \texttt{date\_end}: the date when the process of collecting ballots ended.
            \item \texttt{language}: the language of the descriptions of the projects (i.e., full names of the projects)
            \item \texttt{edition}
            \item \texttt{district}
            \item \texttt{comment}
            \item if \texttt{vote\_type} = \texttt{approval}:
                \begin{itemize}
                    \item \texttt{min\_length} [default: 1]
                    \item \texttt{max\_length} [default: num\_projects]
                    \item \texttt{min\_sum\_cost} [default: 0]
                    \item \texttt{max\_sum\_cost} [default: $\infty$]
                \end{itemize}
            \item if \texttt{vote\_type} = \texttt{ordinal}:
                \begin{itemize}
                    \item \texttt{min\_length} [default: 1]
                    \item \texttt{max\_length} [default: num\_projects]
                    \item \texttt{scoring\_fn} [default: Borda]
                \end{itemize}

            \item if \texttt{vote\_type} = \texttt{cumulative}:
                \begin{itemize}
                    \item \texttt{min\_length} [default: 1]
                    \item \texttt{max\_length} [default: num\_projects]
                    \item \texttt{min\_points} [default: 0]
                    \item \texttt{max\_points} [default: max\_sum\_points]
                    \item \texttt{min\_sum\_points} [default: 0]
                    \item \bftt{max\_sum\_points} 
                \end{itemize}
            \item if \texttt{vote\_type} = \texttt{scoring}:
                \begin{itemize}                    
                    \item \texttt{min\_length} [default: 1]
                    \item \texttt{max\_length} [default: num\_projects]
                    \item \texttt{min\_points} [default: $-\infty$]
                    \item \texttt{max\_points} [default: $\infty$]
                    \item \texttt{default\_score} [default: 0]
                \end{itemize}
            \item \texttt{non-standard fields}
        \end{itemize}
        \item \bftt{value}: the value of the corresponding field.
    \end{itemize}

\subsubsection{Section 2: PROJECTS}
        
    \begin{itemize}
        \item \bftt{project\_id}
        \item \bftt{cost}
        \item \texttt{name}: the full name of the project.
        \item \texttt{category}: the list of tags depscribing the project, separated with commas; for example: education, sport, health, culture, environmental protection, public space, public transit, roads.
        \item \texttt{target}: type voters that might be especially interested in the project.  For example: adults, seniors, children, youth, people with disabilities, families with children, animals.
        \item \texttt{non-standard fields}
    \end{itemize}

\subsubsection{Section 3: VOTES}
    \begin{itemize}
        \item \bftt{voter\_id}
        \item \texttt{age}
        \item \texttt{sex}
        \item \texttt{voting\_method} (e.g., paper, Internet, mail)
        \item if \texttt{vote\_type} = \texttt{approval}:
            \begin{itemize}
                    \item \bftt{vote}: identifiers of the approved projects, separated with commas.
            \end{itemize}
        \item if \texttt{vote\_type} = \texttt{ordinal}:
            \begin{itemize}
                    \item \bftt{vote}: identifiers of the selected projects, from the most preferred to the least preferred one, separated with commas.
            \end{itemize}
        \item if \texttt{vote\_type} = \texttt{cumulative}:
            \begin{itemize}
                \item \bftt{vote}: identifiers of the projects, separated with commas; projects not listed are assumed to get $0$ points. Projects are listed in the decreasing order of the number of points they got from the voter,
                \item \bftt{points}: points given to the projects, listed in the same order as the project identifiers in the field \bftt{vote}.
            \end{itemize}
        \item if \texttt{vote\_type} = \texttt{scoring}:
                    \begin{itemize}
                    \item \bftt{vote}: project identifiers, separated with commas; projects not listed are assumed to get \texttt{default\_score} points. Projects are listed in the decreasing order of the number of points they got from the voter.
                    \item \bftt{points}: points given to the projects, listed in the same order as the project identifiers in the field \bftt{vote}.
            \end{itemize}
            
        \item \texttt{non-standard fields}
        
       \end{itemize}



\newpage

\vspace{-3mm}
\section{Additional figures}\label{sec:extendedplots}
\vspace{-3mm}

\begin{figure*}[!h]
    \centering
    \begin{subfigure}[t]{0.19\textwidth}
        \includegraphics[width=0.9\textwidth]{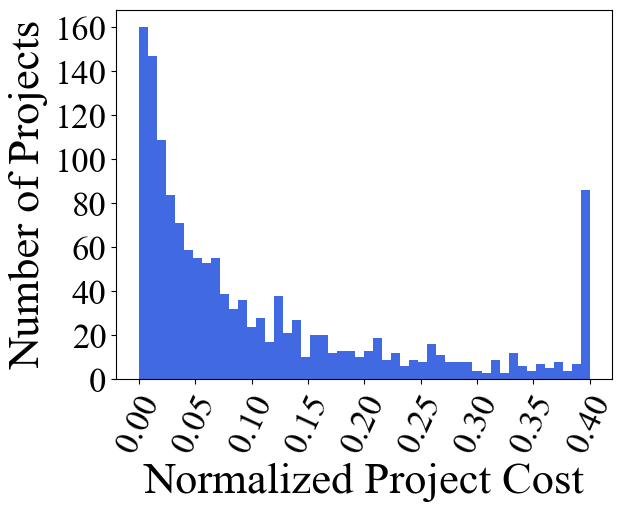}
        \caption{Krakow 2020-2022}
    \end{subfigure}%
    \begin{subfigure}[t]{0.19\textwidth}
        \includegraphics[width=0.9\textwidth]{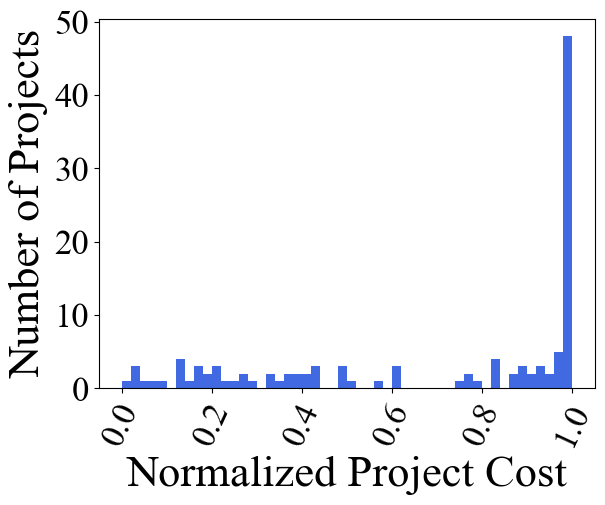}
        \caption{Zabrze 2020-2021}
    \end{subfigure}%
    \begin{subfigure}[t]{0.19\textwidth}
        \includegraphics[width=0.9\textwidth]{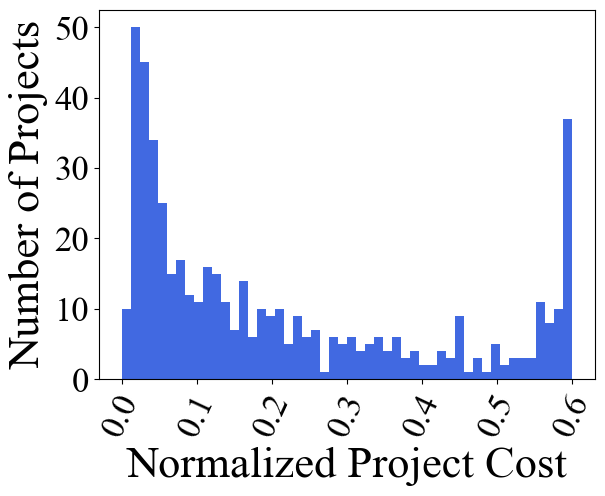}
        \caption{Katowice 2020-2021}
    \end{subfigure}%
    \begin{subfigure}[t]{0.19\textwidth}
        \includegraphics[width=0.9\textwidth]{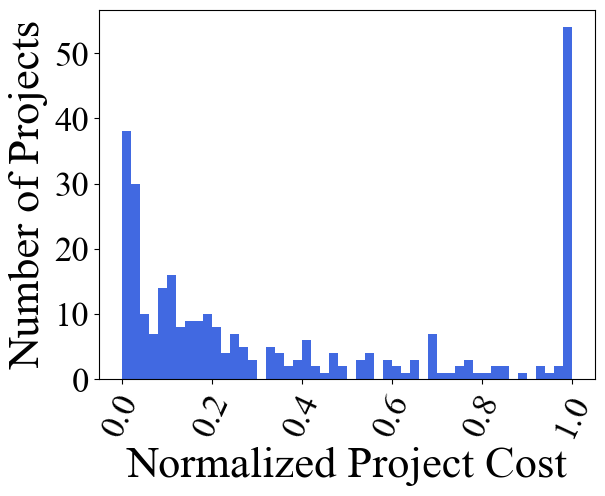}
        \caption{Gdansk 2020}
    \end{subfigure}%
    \begin{subfigure}[t]{0.19\textwidth}
        \includegraphics[width=0.9\textwidth]{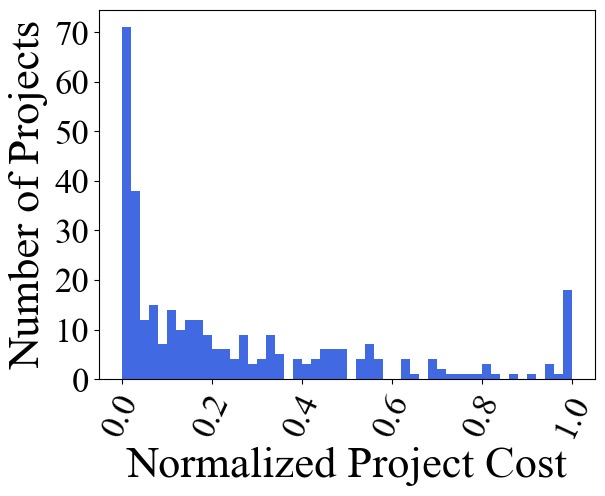}
        \caption{Czestochowa 2020}
    \end{subfigure}

\begin{subfigure}[t]{0.19\textwidth}
        \includegraphics[width=0.9\textwidth]{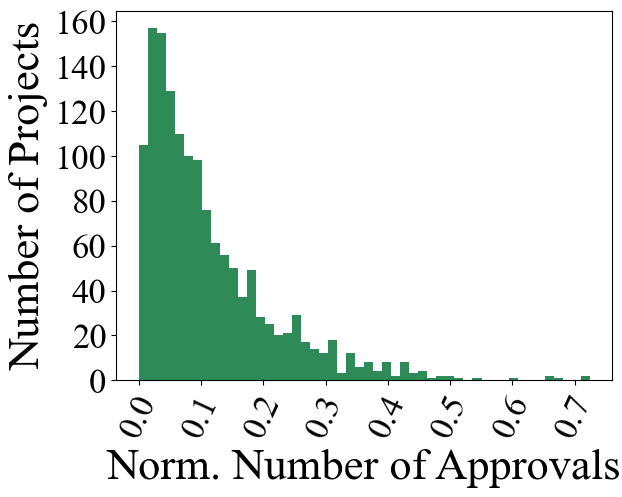}
        \caption{Krakow 2020-2022}
    \end{subfigure}%
    \begin{subfigure}[t]{0.19\textwidth}
        \includegraphics[width=0.9\textwidth]{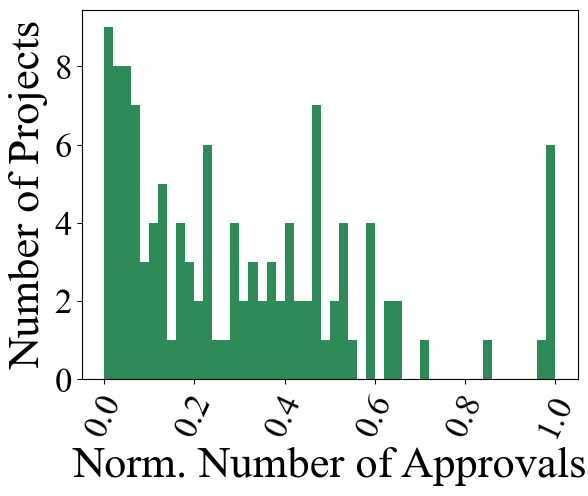}
        \caption{Zabrze 2020-2021}
    \end{subfigure}%
    \begin{subfigure}[t]{0.19\textwidth}
        \includegraphics[width=0.9\textwidth]{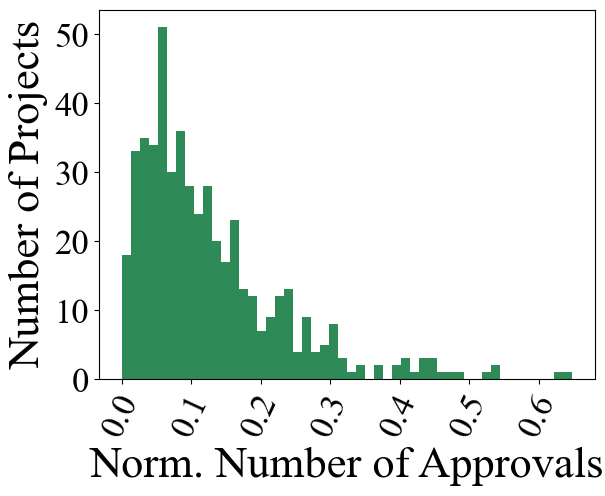}
        \caption{Katowice 2020-2021}
    \end{subfigure}%
    \begin{subfigure}[t]{0.19\textwidth}
        \includegraphics[width=0.9\textwidth]{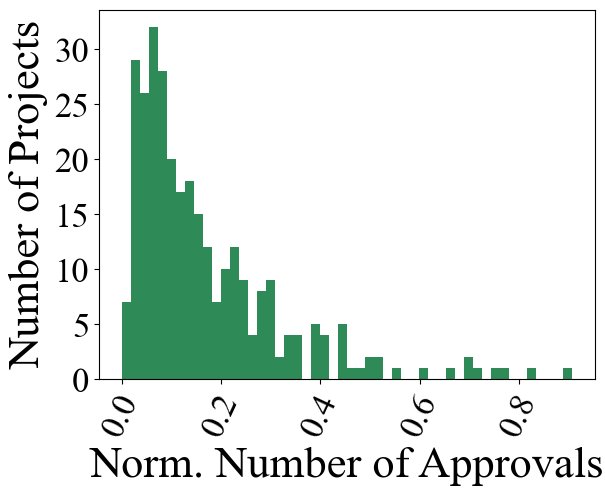}
        \caption{Gdansk 2020}
    \end{subfigure}%
    \begin{subfigure}[t]{0.19\textwidth}
        \includegraphics[width=0.9\textwidth]{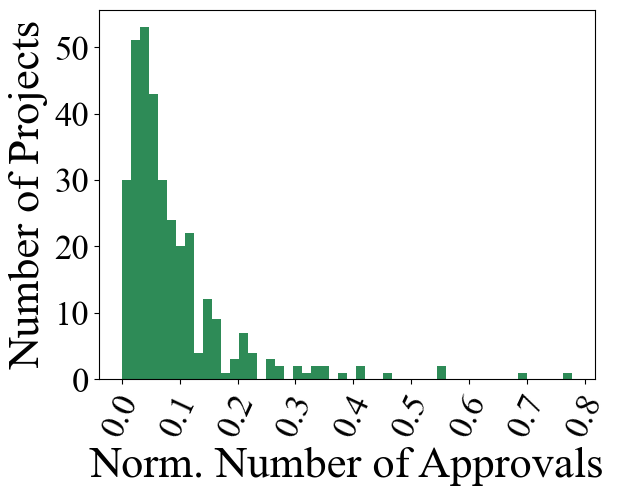}
        \caption{Czestochowa 2020}
    \end{subfigure}

\begin{subfigure}[t]{0.19\textwidth}
        \includegraphics[width=0.9\textwidth]{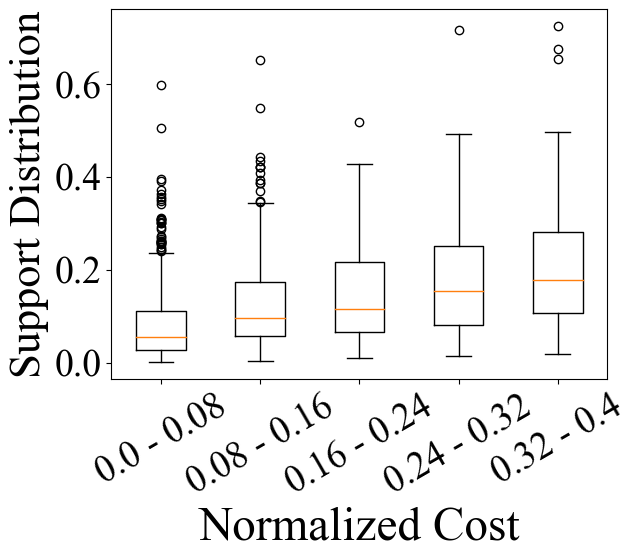}
        \caption{Krakow 2020-2022}
    \end{subfigure}%
    \begin{subfigure}[t]{0.19\textwidth}
        \includegraphics[width=0.9\textwidth]{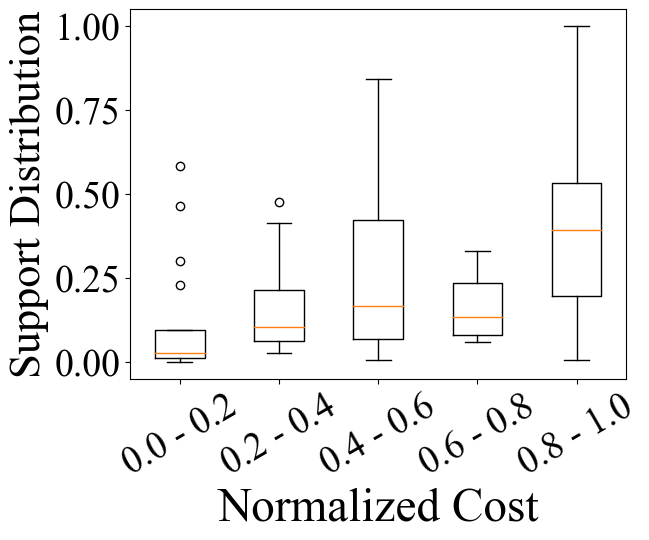}
        \caption{Zabrze 2020-2021}
    \end{subfigure}%
    \begin{subfigure}[t]{0.19\textwidth}
        \includegraphics[width=0.9\textwidth]{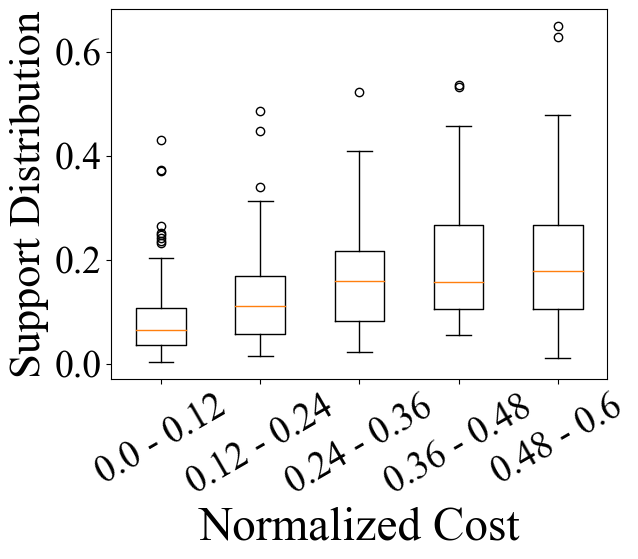}
        \caption{Katowice 2020-2021}
    \end{subfigure}%
    \begin{subfigure}[t]{0.19\textwidth}
        \includegraphics[width=0.9\textwidth]{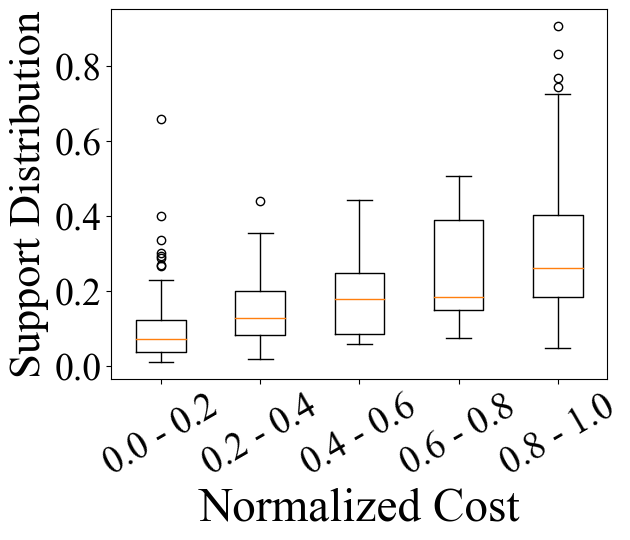}
        \caption{Gdansk 2020}
    \end{subfigure}%
    \begin{subfigure}[t]{0.19\textwidth}
        \includegraphics[width=0.9\textwidth]{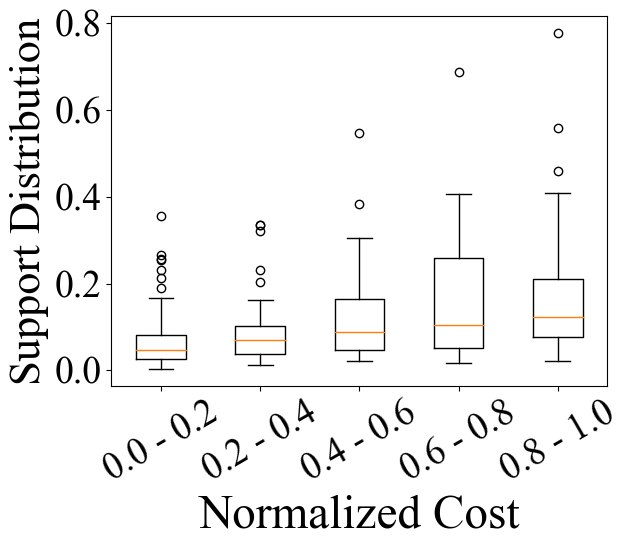}
        \caption{Czestochowa 2020}
    \end{subfigure}

\begin{subfigure}[t]{0.19\textwidth}
        \includegraphics[width=0.9\textwidth]{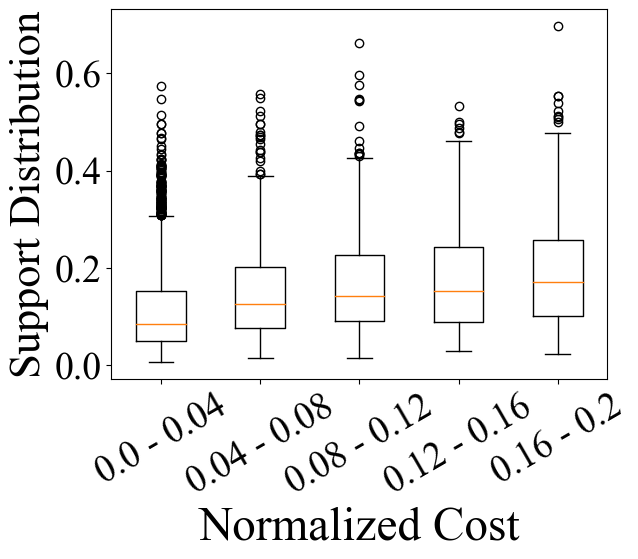}
        \caption{Warsaw 2020-2023}
    \end{subfigure}%
    \begin{subfigure}[t]{0.19\textwidth}
        \includegraphics[width=0.9\textwidth]{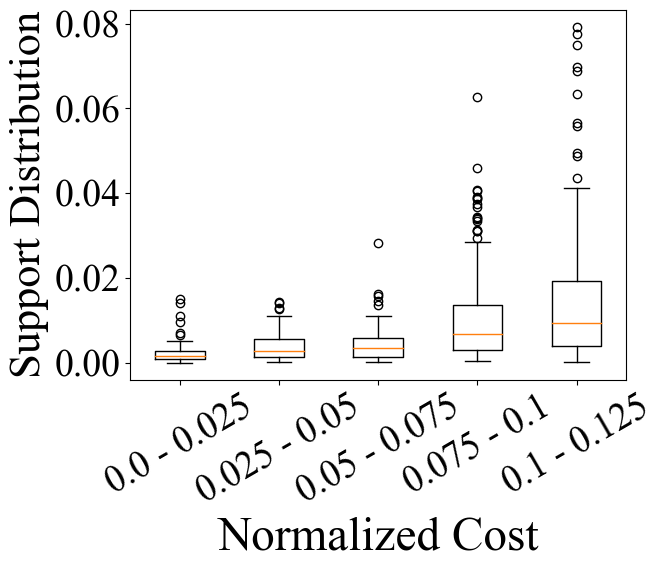}
        \caption{Wroclaw 2019-2021}
\end{subfigure}%

    \caption{\label{fig:project_cost_and_support_apdx} The distribution of costs and the total number of approvals per project. The histograms contain aggregated data from different years. For each PB instance we normalize the costs by dividing them by the total budget, and the number of approvals by dividing it by the number of voters in a given instance.}
\end{figure*}

\begin{figure}[!h]
\begin{center}
\includegraphics[width=\textwidth]{plots/legend.pdf}
\begin{subfigure}[t]{0.5\textwidth}
  \includegraphics[width=\textwidth]{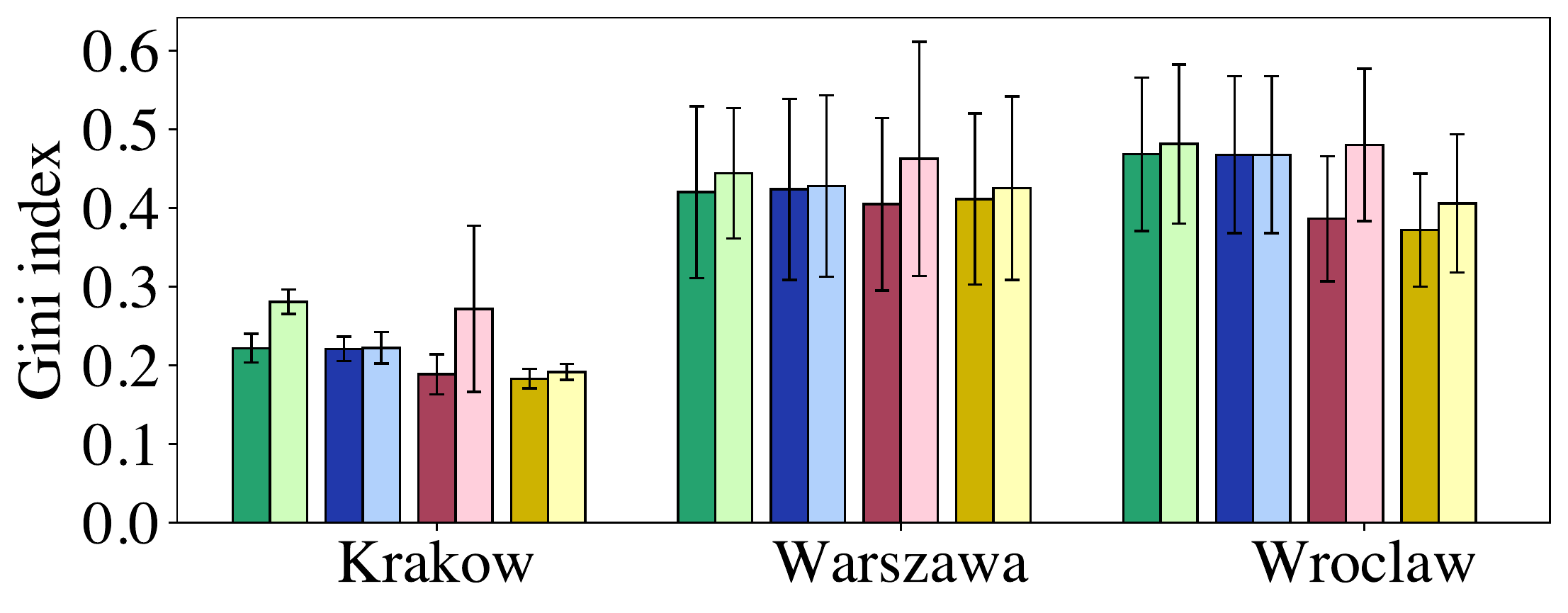}
  \end{subfigure}
  \end{center}
  \caption{Comparison of Equal Shares and Utilitarian Greedy with respect to the Gini index.
  The label ``Cost'' means that we are referring to the cost-utility variant of the method; otherwise we are referring to its score-utility variant. The symbols ``D'' and ``C'' stand for the districtwise and citywide schemes, respectively.}
  \label{fig:extended_pops_es_vs_ug}
\end{figure}

\begin{figure}[!h]
\includegraphics[width=\linewidth]{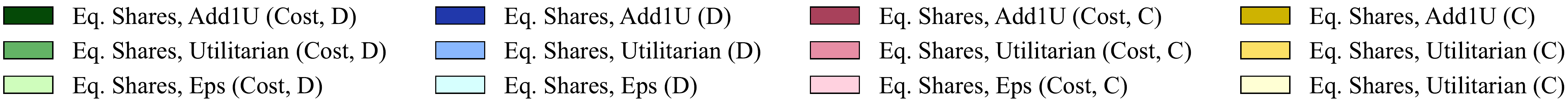}

\begin{subfigure}[t]{0.49\textwidth}
  \includegraphics[width=\linewidth]{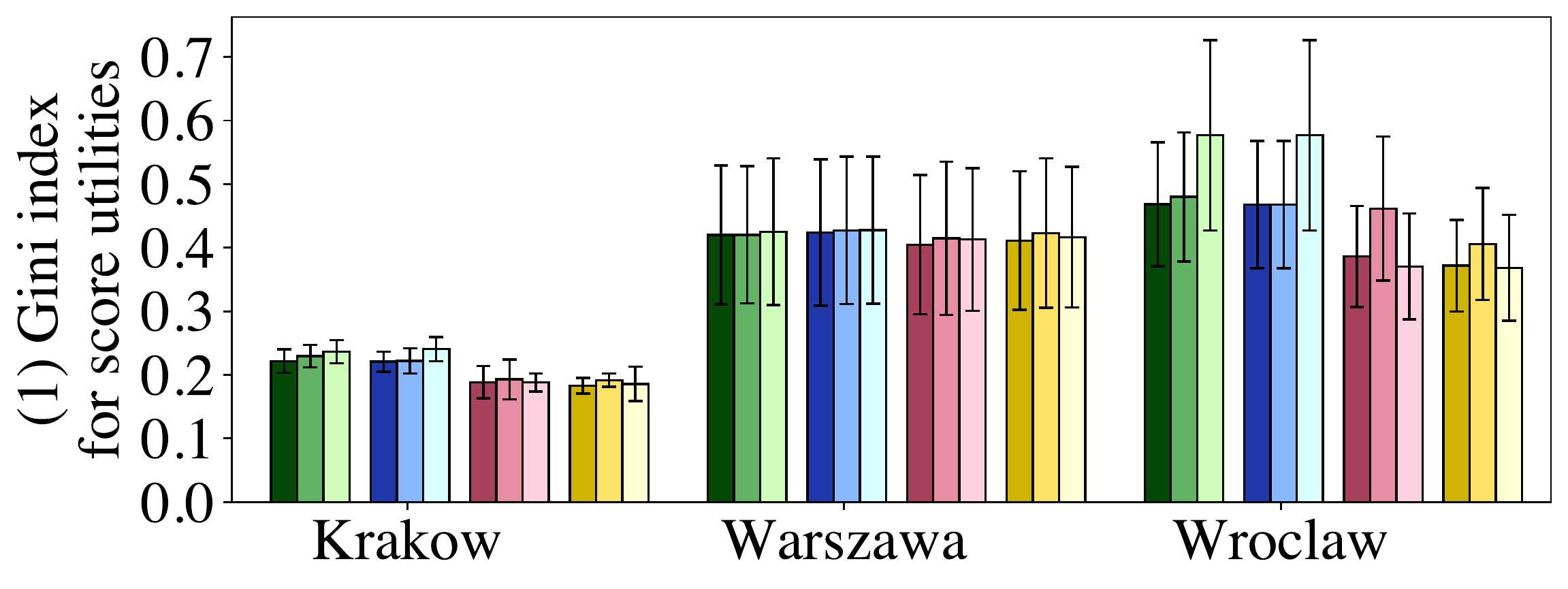}
\end{subfigure}
\begin{subfigure}[t]{0.49\textwidth}
  \includegraphics[width=\linewidth]{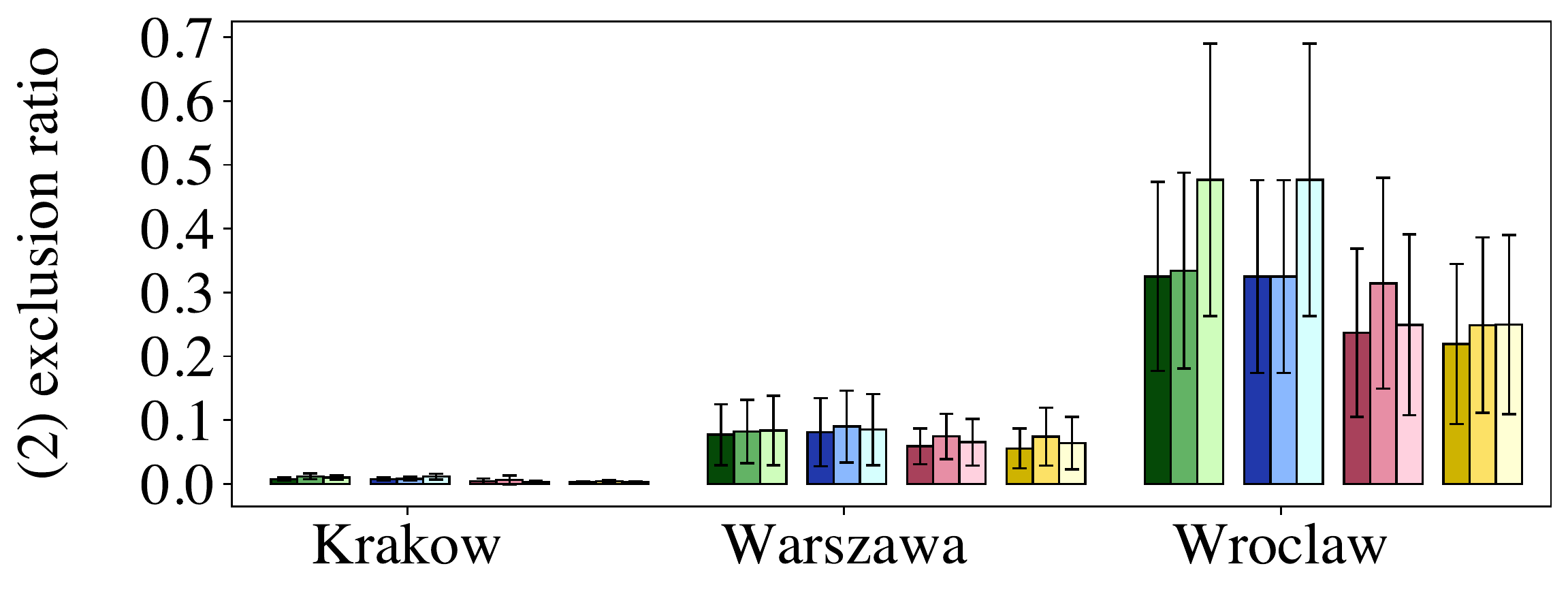}
\end{subfigure}
  \caption{Comparison of different completion rules for the Method of Equal Shares. We compare (1) the Gini index of the score utility, and (2) the exclusion ratio. The label ``Cost'' means that we are referring to the cost-utility variant of the method; otherwise we are referring to its score-utility variant. The symbols ``D'' and ``C'' stand for the districtwise and citywide schemes, respectively.}\label{fig:extended_pops_completions}
\end{figure}

\begin{figure}[!h]
  \centering
  \includegraphics[width=0.5\linewidth]{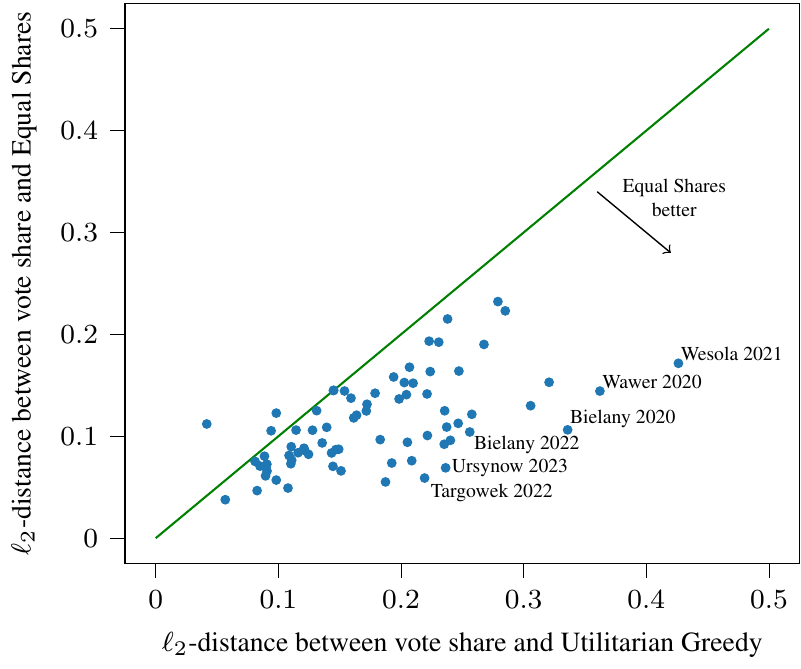}
  \caption{Comparison of Equal Shares (cost utility, Add1U) and Utilitarian Greedy with respect to how well the voters' preferences across categories are reflected by the spending of the rules. Each Warsaw district in each year 2020--23 is represented by a blue point, placed according to the $\ell_2$-distance between the vote share vector and the spending shares for the two rules. For points below the green line, Utilitarian Greedy has a higher $\ell_2$-distance. Points with a particularly large imbalance are labelled.
  }\label{fig:division_into_categories}
\end{figure}

\clearpage
\section{Additional Maps of Elections}\label{sec:maps_of_elections_apdx}





\begin{figure*}[!h]
    \centering
    \includegraphics[width=15cm]{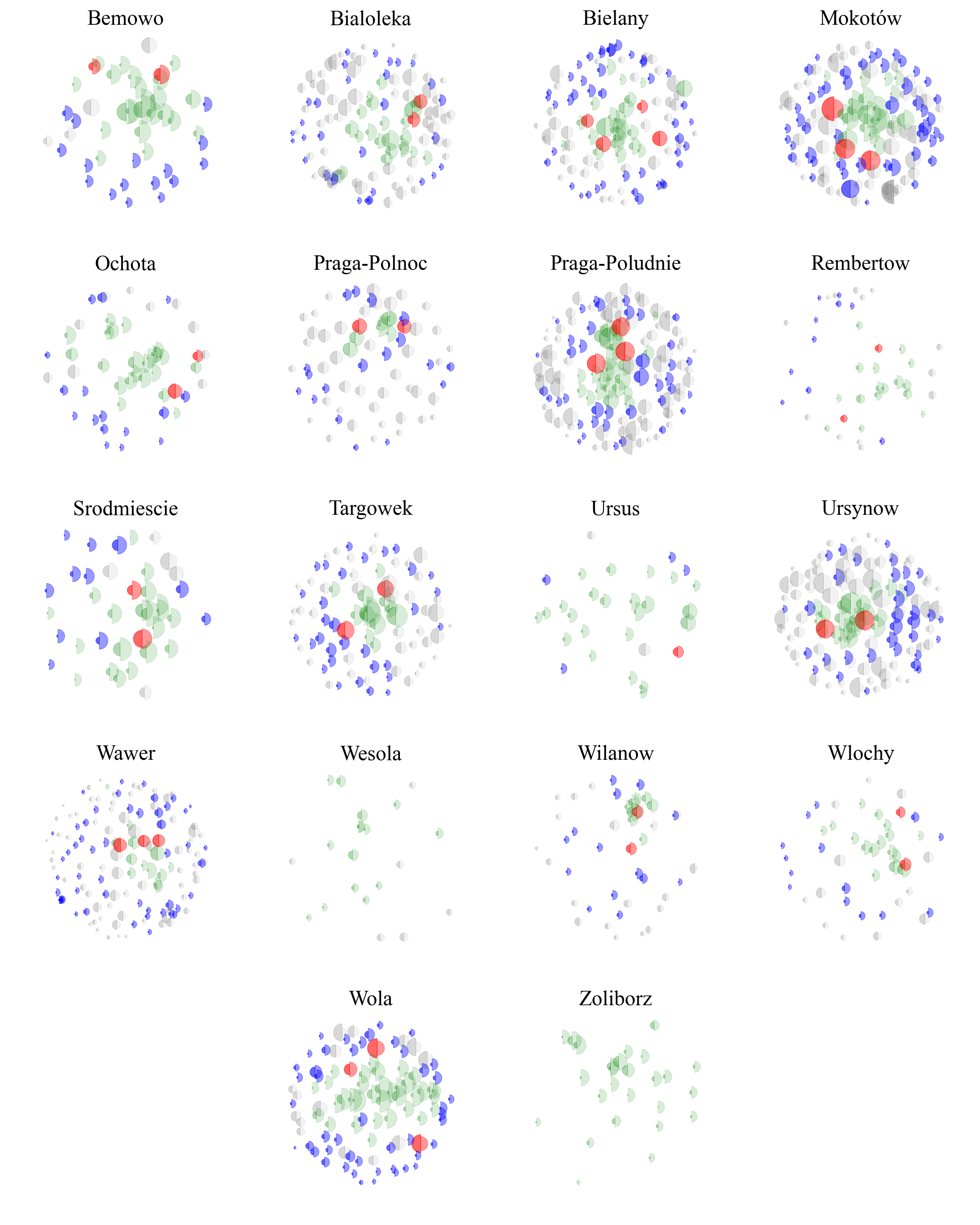}
    \caption{Visualisation of projects in PB elections from \textbf{Warsaw 2020 using the Jaccard distance}. Each 
    project is represented by two glued-together half-discs. The size of the left half    
    is proportional to the project's cost, whereas the size of the right half is proportional to the total number of votes the project received. The figures compare the outcomes of the cost-utility variant of Equal Shares with Add1U completion, with the outcomes of the Utilitarian Greedy rule.
    Specifically, gray projects were not selected by either of the rules,  green projects were selected by both, blue projects were selected only by Equal Shares, and red projects were selected only by Utilitarian Greedy.}
\end{figure*}

\begin{figure*}[h!]
    \centering
    \includegraphics[width=16cm]{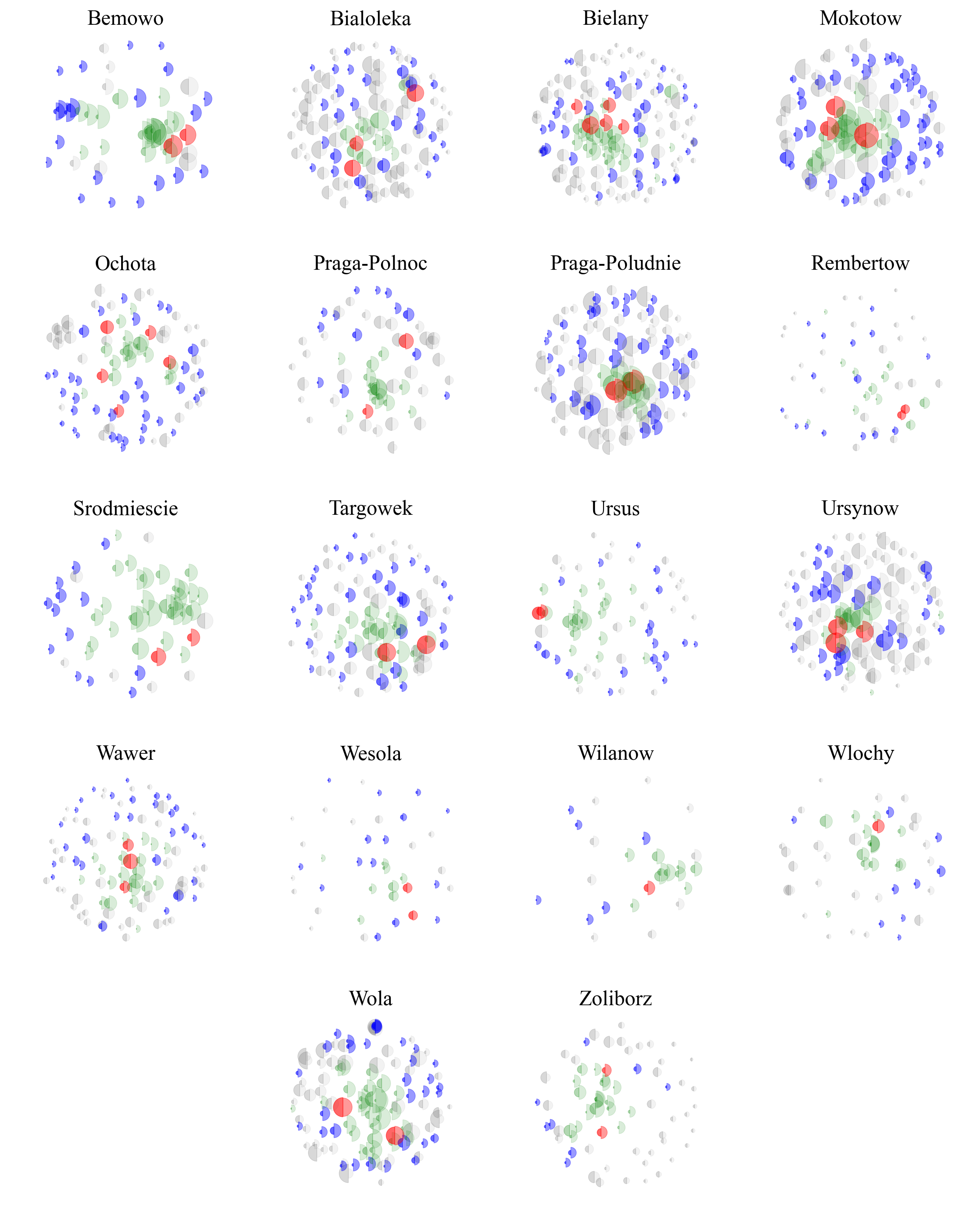}
    \caption{Visualisation of projects in PB elections from \textbf{Warsaw 2021 using the Jaccard distance}. Each 
    project is represented by two glued-together half-discs. The size of the left half    
    is proportional to the project's cost, whereas the size of the right half is proportional to the total number of votes the project received. The figures compare the outcomes of the cost-utility variant of Equal Shares with Add1U completion, with the outcomes of the Utilitarian Greedy rule.
    Specifically, gray projects were not selected by either of the rules,  green projects were selected by both, blue projects were selected only by Equal Shares, and red projects were selected only by Utilitarian Greedy.}
\end{figure*}

\begin{figure*}[t]
    \centering
    \includegraphics[width=16cm]{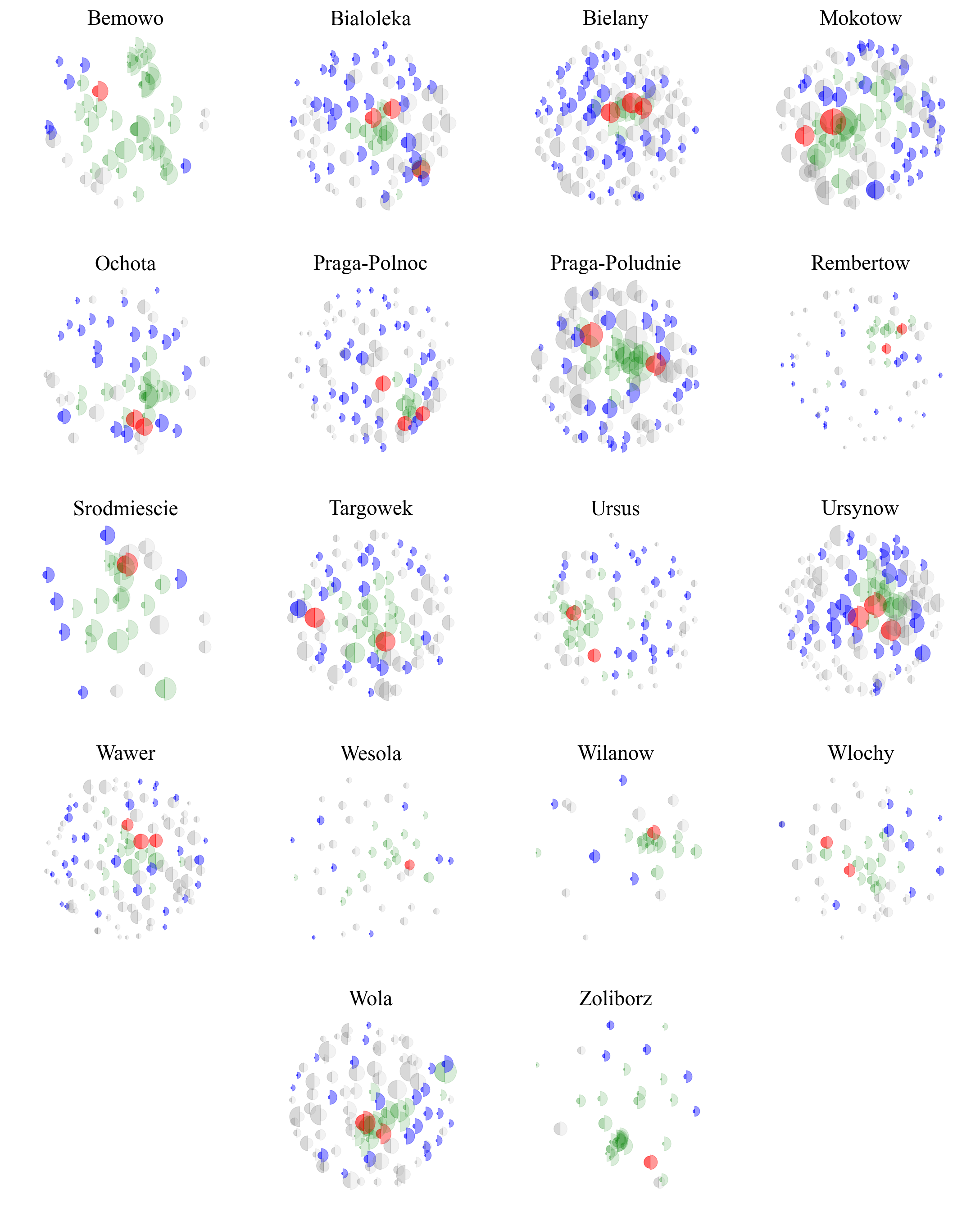}
    \caption{Visualisation of projects in PB elections from \textbf{Warsaw 2022 using the Jaccard distance}. Each 
    project is represented by two glued-together half-discs. The size of the left half    
    is proportional to the project's cost, whereas the size of the right half is proportional to the total number of votes the project received. The figures compare the outcomes of the cost-utility variant of Equal Shares with Add1U completion, with the outcomes of the Utilitarian Greedy rule.
    Specifically, gray projects were not selected by either of the rules,  green projects were selected by both, blue projects were selected only by Equal Shares, and red projects were selected only by Utilitarian Greedy.}
\end{figure*}

\begin{figure*}[h!]
    \centering
    \includegraphics[width=16cm]{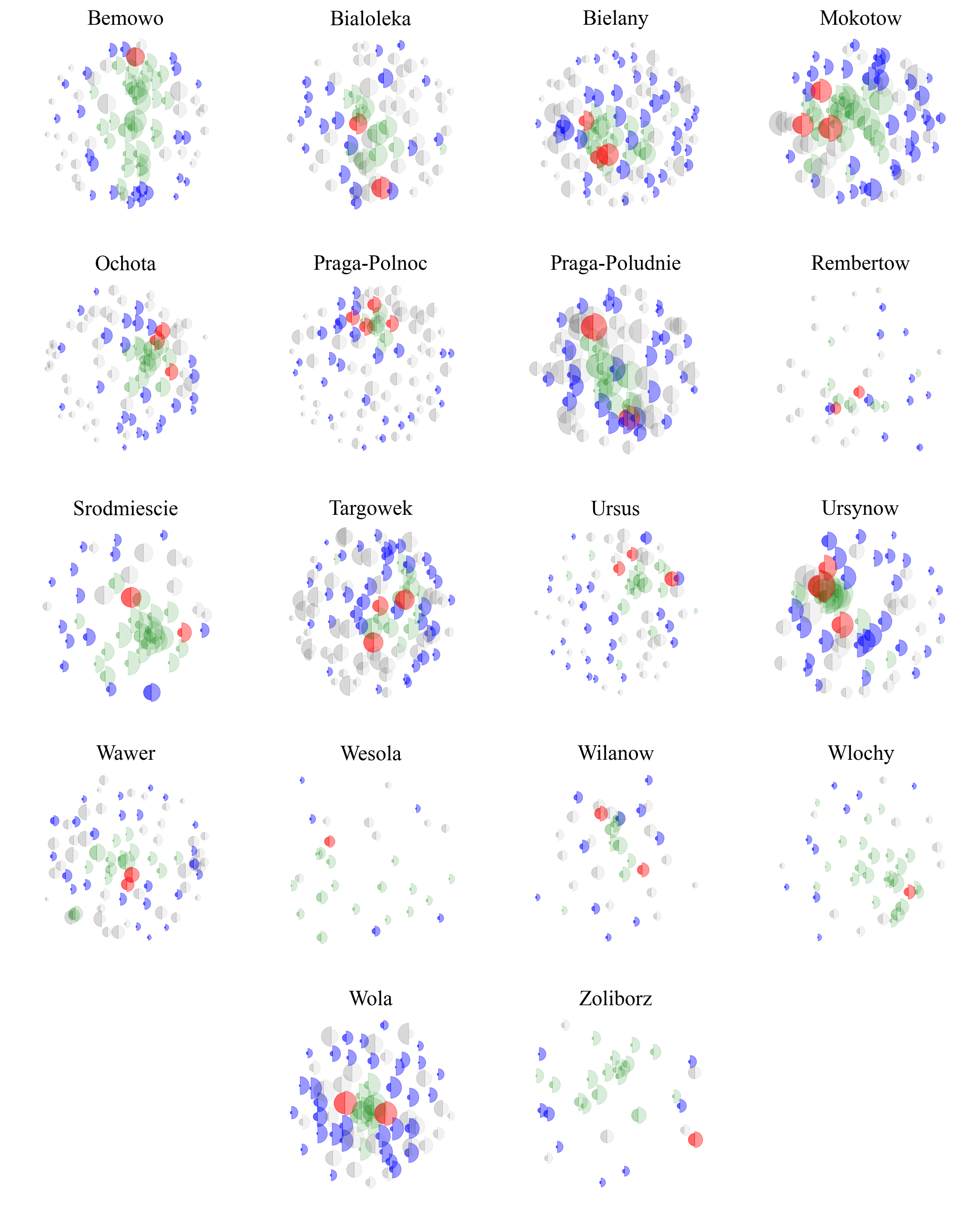}
    \caption{Visualisation of projects in PB elections from \textbf{Warsaw 2023 using the Jaccard distance}. Each 
    project is represented by two glued-together half-discs. The size of the left half    
    is proportional to the project's cost, whereas the size of the right half is proportional to the total number of votes the project received. The figures compare the outcomes of the cost-utility variant of Equal Shares with Add1U completion, with the outcomes of the Utilitarian Greedy rule.
    Specifically, gray projects were not selected by either of the rules,  green projects were selected by both, blue projects were selected only by Equal Shares, and red projects were selected only by Utilitarian Greedy.}
\end{figure*}

\begin{figure*}[h!]
    \centering
    \includegraphics[width=16cm]{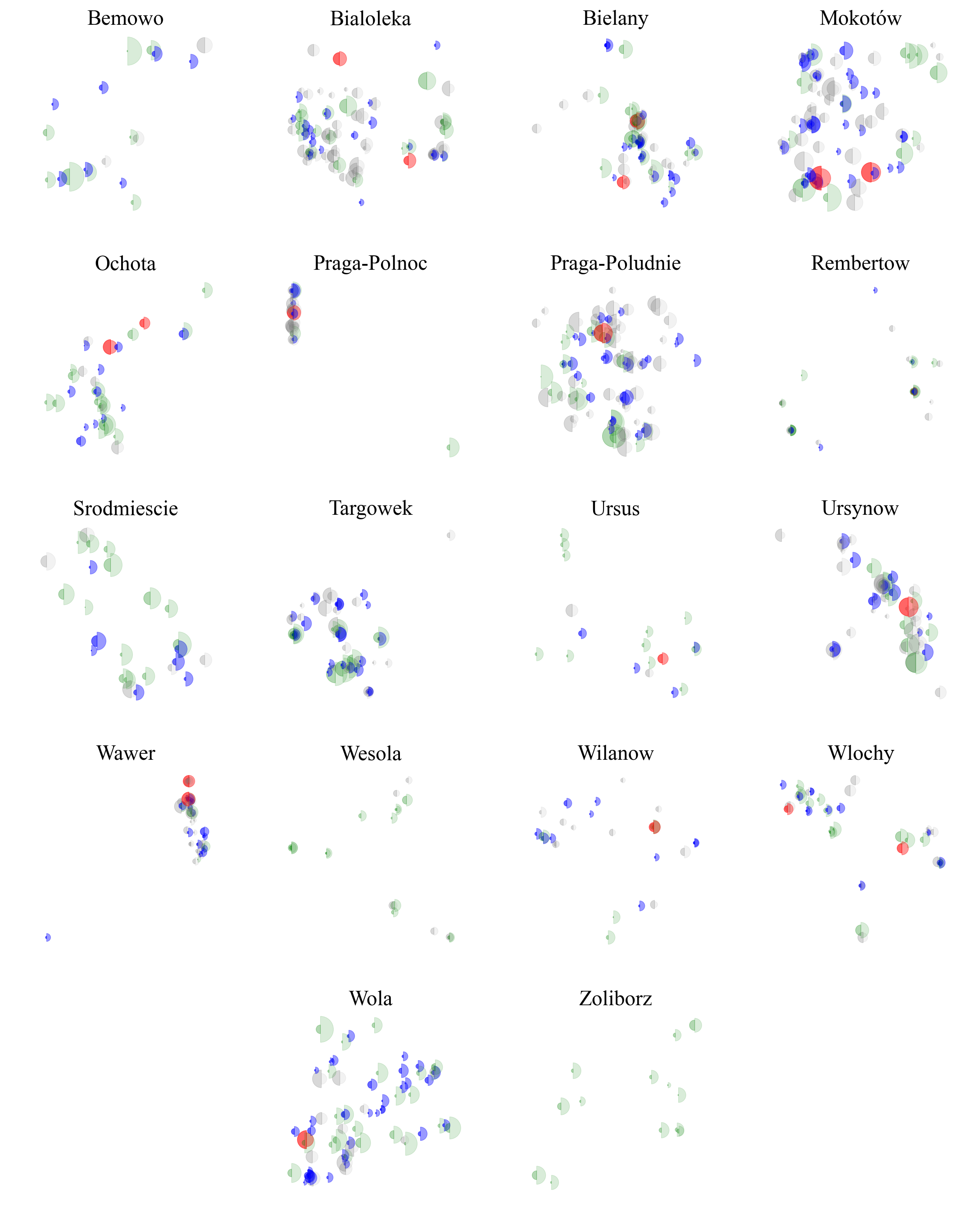}
    \caption{Visualisation of projects in PB elections from \textbf{Warsaw 2020 using the GPS data}. Each 
    project is represented by two glued-together half-discs. The size of the left half    
    is proportional to the project's cost, whereas the size of the right half is proportional to the total number of votes the project received. The figures compare the outcomes of the cost-utility variant of Equal Shares with Add1U completion, with the outcomes of the Utilitarian Greedy rule.
    Specifically, gray projects were not selected by either of the rules,  green projects were selected by both, blue projects were selected only by Equal Shares, and red projects were selected only by Utilitarian Greedy.}
\end{figure*}

\begin{figure*}[h!]
    \centering
    \includegraphics[width=16cm]{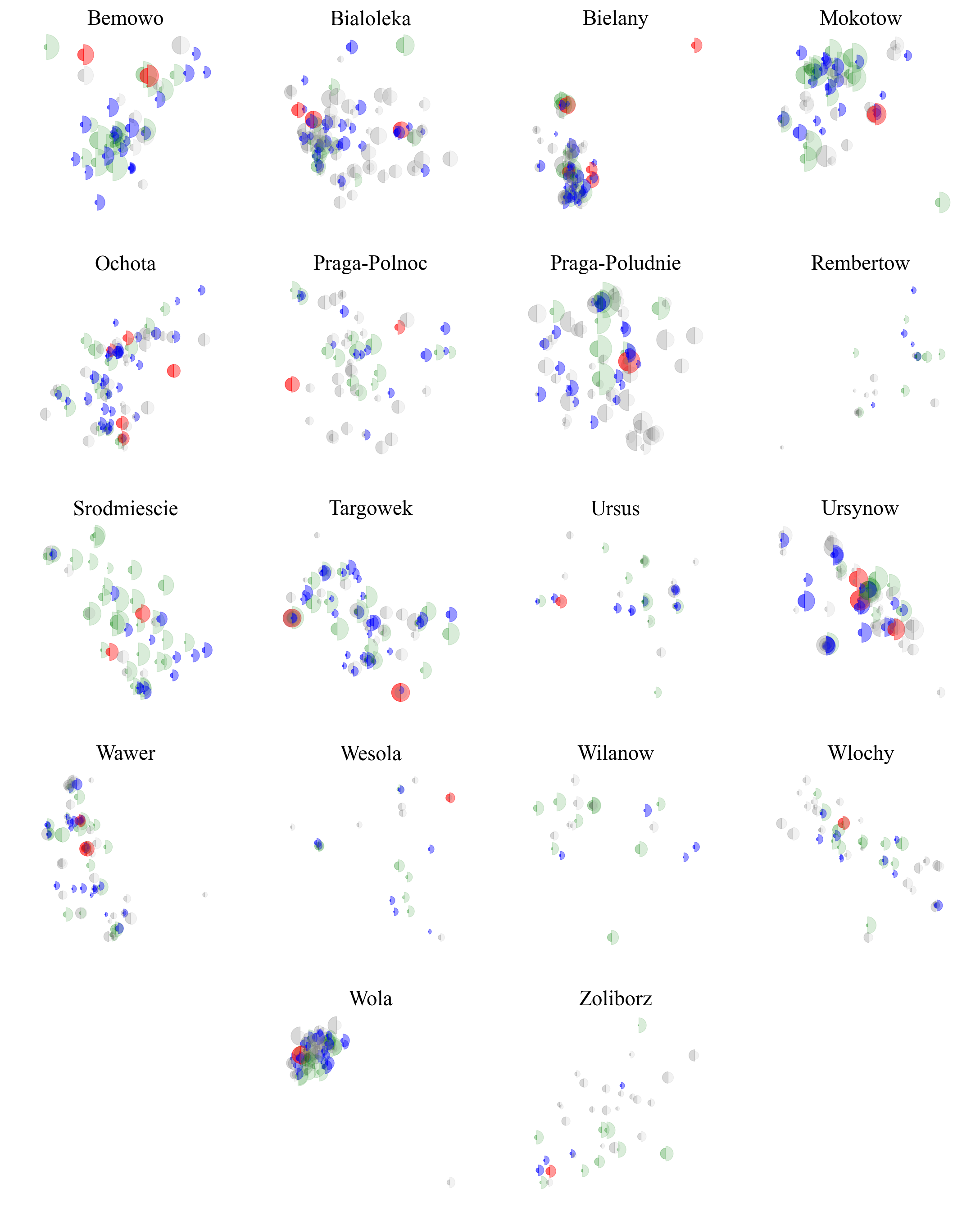}
    \caption{Visualisation of projects in PB elections from \textbf{Warsaw 2021 using the GPS data}. Each 
    project is represented by two glued-together half-discs. The size of the left half    
    is proportional to the project's cost, whereas the size of the right half is proportional to the total number of votes the project received. The figures compare the outcomes of the cost-utility variant of Equal Shares with Add1U completion, with the outcomes of the Utilitarian Greedy rule.
    Specifically, gray projects were not selected by either of the rules,  green projects were selected by both, blue projects were selected only by Equal Shares, and red projects were selected only by Utilitarian Greedy.}
\end{figure*}

\begin{figure*}[h!]
    \centering
    \includegraphics[width=16cm]{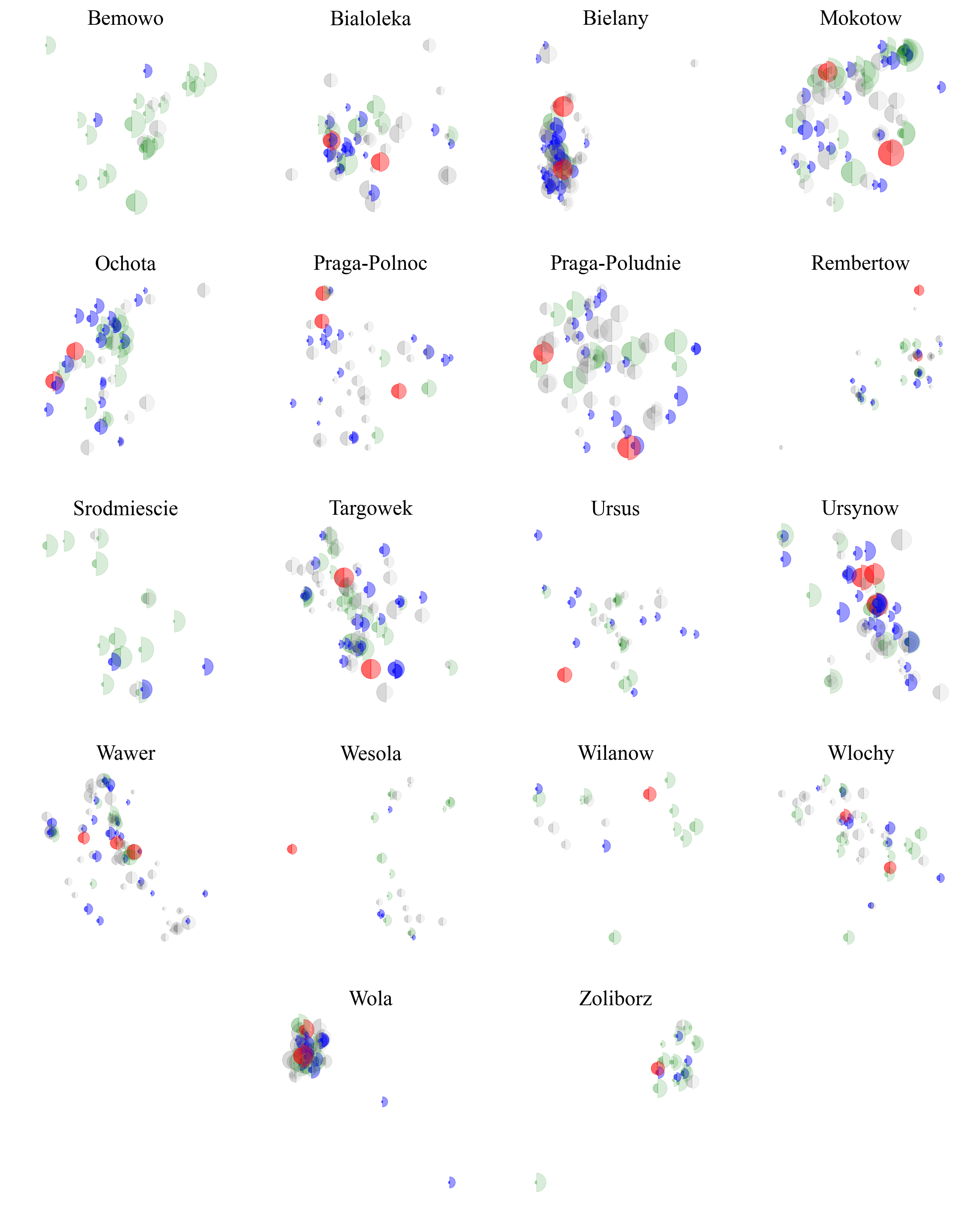}
    \caption{Visualisation of projects in PB elections from \textbf{Warsaw 2022 using the GPS data}. Each 
    project is represented by two glued-together half-discs. The size of the left half    
    is proportional to the project's cost, whereas the size of the right half is proportional to the total number of votes the project received. The figures compare the outcomes of the cost-utility variant of Equal Shares with Add1U completion, with the outcomes of the Utilitarian Greedy rule.
    Specifically, gray projects were not selected by either of the rules,  green projects were selected by both, blue projects were selected only by Equal Shares, and red projects were selected only by Utilitarian Greedy.}
\end{figure*}

\begin{figure*}[h!]
    \centering
    \includegraphics[width=16cm]{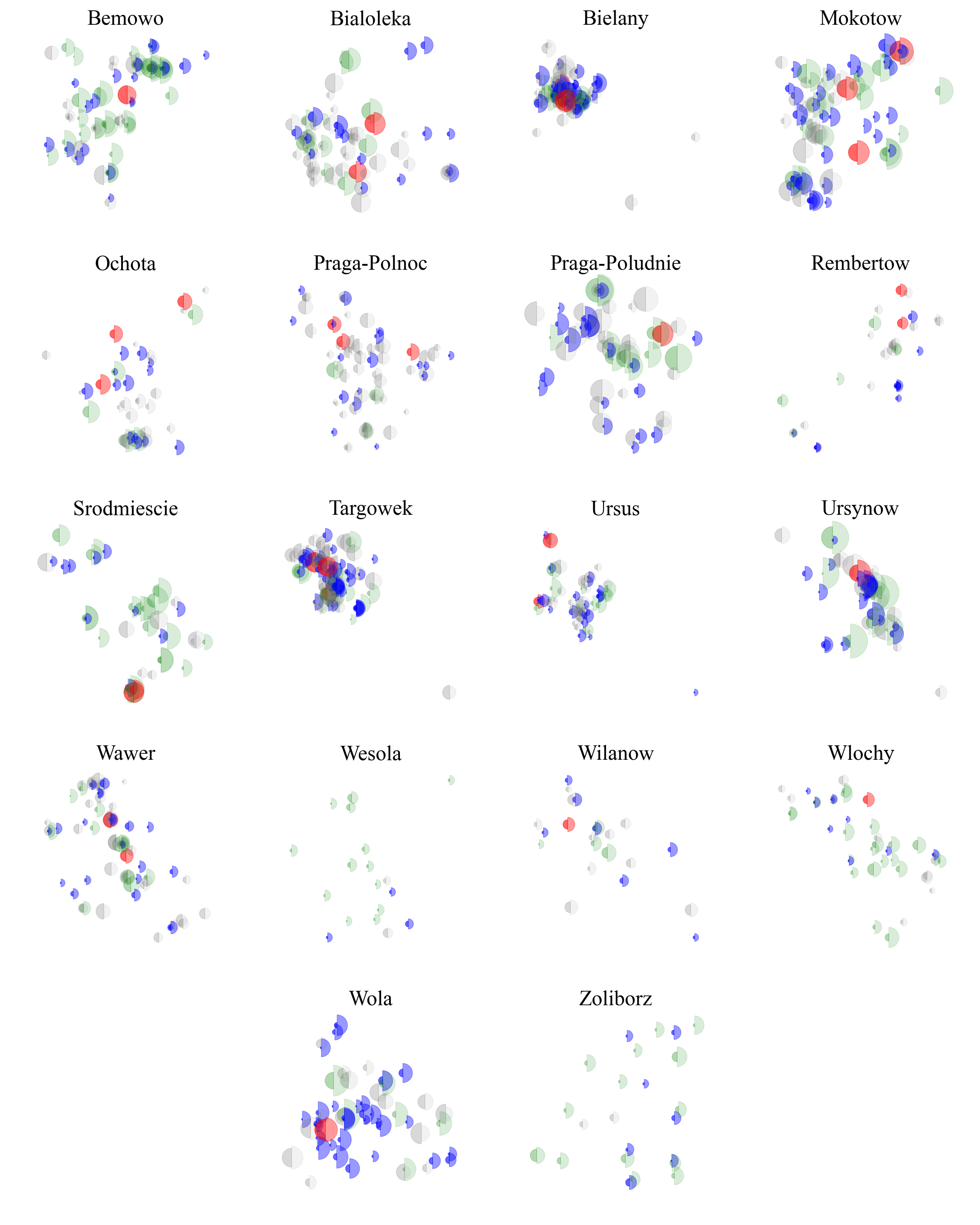}
    \caption{Visualisation of projects in PB elections from \textbf{Warsaw 2023 using the GPS data}. Each 
    project is represented by two glued-together half-discs. The size of the left half    
    is proportional to the project's cost, whereas the size of the right half is proportional to the total number of votes the project received. The figures compare the outcomes of the cost-utility variant of Equal Shares with Add1U completion, with the outcomes of the Utilitarian Greedy rule.
    Specifically, gray projects were not selected by either of the rules,  green projects were selected by both, blue projects were selected only by Equal Shares, and red projects were selected only by Utilitarian Greedy.}
\end{figure*}

\end{document}